%% file: main.tex
\documentclass[preprint]{elsarticle}
\usepackage{lineno,hyperref}
\usepackage{framed,multirow}
\usepackage{subfigure}
\usepackage{graphicx}
\usepackage{multicol}

\usepackage{amssymb}  
\usepackage{amsthm}
\usepackage[cmex10]{amsmath}
\usepackage{latexsym}

\usepackage{booktabs}
\usepackage{longtable}
\usepackage{tabularx}
\usepackage{multirow}

\newcolumntype{C}{>{\centering\arraybackslash}X}
\newcolumntype{L}{>{\raggedright\arraybackslash}X}
\newcolumntype{R}{>{\raggedleft\arraybackslash}X}
\newcolumntype{P}[1]{>{\raggedright\arraybackslash}p{#1}}

\usepackage{url}
\usepackage{xcolor}

\usepackage{hyperref}

\definecolor{newcolor}{rgb}{.8,.349,.1}

\newcommand{\ceil}[1]{\lceil #1 \rceil}
\newcommand{\floor}[1]{\lfloor #1 \rfloor}
\def\sq#1{\lq{#1}\rq}

\def\abss#1{\lvert \lvert {#1} \rvert \rvert}

\usepackage[left=4cm,right=4cm]{geometry}

\journal{arXiv}

\begin{document}

\begin{frontmatter}

\title{\bf{Deep motion estimation for parallel inter-frame prediction in video compression}}

\author{Andr\'e Nortje}
\ead{18247717@sun.ac.za}
\author{Herman A. Engelbrecht}
\ead{hebrecht@sun.ac.za}
\author{Herman Kamper}
\ead{kamperh@sun.ac.za}

\address{Department of Electrical and Electronic Engineering,\\ 
Stellenbosch University, South Africa}

\input{my_content/abstract.tex}


\end{frontmatter}

\input{my_content/intro.tex}

\input{my_content/arch.tex}

\input{my_content/train.tex}

\input{my_content/exp.tex}

\input{my_content/eval.tex}

\input{my_content/conc.tex}

\bibliographystyle{model1-num-names.bst}
\bibliography{my_refs}

\end{document}

%% file: my_content/abstract.tex

\begin{abstract}
Standard video codecs rely on optical flow to guide inter-frame prediction: pixels from reference frames are moved via motion vectors to predict target video frames. We propose to learn binary motion codes that are encoded based on an input video sequence. These codes are not limited to 2D translations, but can capture complex motion (warping, rotation and occlusion). Our motion codes are learned as part of a single neural network which also learns to compress and decode them. This approach supports parallel video frame decoding instead of the sequential motion estimation and compensation of flow-based methods. We also introduce 3D dynamic bit assignment to adapt to object displacements caused by motion, yielding additional bit savings. By replacing the optical flow-based block-motion algorithms found in an existing video codec with our learned inter-frame prediction model, our approach outperforms the standard H.264 and H.265 video codecs across at low bitrates.
\end{abstract}

%% file: my_content/intro.tex
%

%
\begin{figure}[!h]
	\center{
		\includegraphics[width=.9\textwidth]{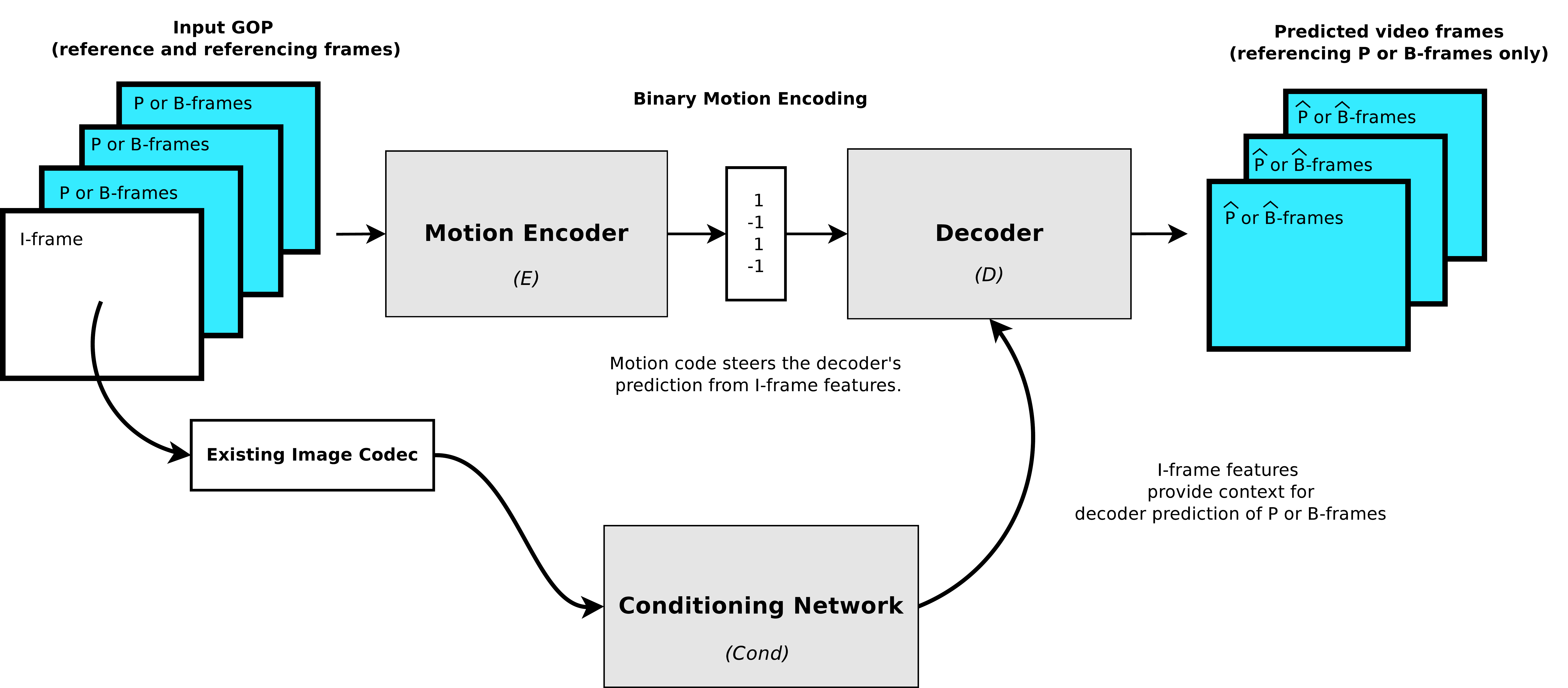}
	}
	\caption{
		Video prediction network architectural overview. 
		A learned binary motion encoding either guides the extrapolation of 
		P-frames (Predicted-frames) from past I-frames (Intra-frame) 
		or the interpolation of B-frames (Bi-directional-frame) from 
		bounding past and future I-frames. 
		Reference I-frames are coded and decoded independently 
		with an existing image codec. 
	}
\label{fig:vid_net}
\end{figure}

\section{Introduction}
Video is predicted to make up 82\% of Internet traffic 
by the year 2022~\cite{Cisco2017}.
Advancing video compression is necessary to curb high bitrates 
and ease bandwidth consumption. 
To this end, learnable deep neural networks are emerging 
as likely successors to hand-tuned standard video codecs~\cite{Rippel2018,Lu2019}.
Inspired by recent work in binary image inpainting~\cite{Nortje2019},
we propose to learn binary motion codes 
for parallel video frame prediction.
Our goal here is to show how this new parallel prediction strategy 
can replace the sequential flow-based methods used by 
existing video codecs to improve their compression efficiency.

Standard video codecs, such as H.264~\cite{ITU-T2003} and H.265~\cite{ITU-T2018}, 
take advantage of the spatial and temporal redundancies in videos to aid compression.
They assign video frames into one of three groups~\cite{Gall1991}:
I-frames, or `intra-frames', are compressed independently from 
surrounding frames by means of an image codec;
P-frames are `predicted-frames' extrapolated from past frames; 
and B-frames are `bi-directionally' interpolated 
from bounding past and future frames.
The compressed I-frames are transmitted directly, 
while the extrapolation and interpolation of P-frames and B-frames 
are achieved via the transmission of 
highly compressible optical flow vectors~\cite{Richards2010}.
These motion vectors (MVs) convey motion by specifying the movement 
of pixels from one frame to another.  

Dense optical flow~\cite{Farneback2003} produces too many MVs
for efficient compression (one per pixel location). 
Consequently, standard video codecs resort to block-based 
motion estimation and compensation techniques~\cite{ITU-T2003,ITU-T2018,Google2016}.
This entails partitioning video frames 
into patches called macroblocks.
In the motion estimation step,
each macroblock in the current frame is related to the location 
of the most similar macroblock in a past or future reference frame 
by means of an MV which contains its displacement in the $x$ and $y$ directions.
Searching for representative macroblocks in the reference frame is computationally 
expensive and numerous search algorithms have been proposed to help 
speed up this process~\cite{Li1994,Po1996,Lu1997,Zhu2000,Nie2002}.
After the transmission of a reference I-frame, 
only MVs need be transmitted to motion-compensate macroblocks in the I-frame
and form predictions of the subsequent frames within a video sequence.
Standard image compression is used to encode the residuals (differences) between 
the vector-based motion predictions and 
the original video frames to improve reconstruction quality.

Block-based motion prediction, although effective, 
is prone to block artefacts and only allows for sequential decoding~\cite{Haskell1997}.
Furthermore, these algorithms suffer from hand-tuned parameterisation 
and lack the ability to undergo joint optimisation with the rest 
of a video compression system.
We present a deep learning approach to video frame prediction that 
can be optimised end-to-end as part of a larger video compression system. 
Given I-frame context, our model is also able to decode 
P-frames or B-frames in parallel without the additional overhead 
of motion-estimation search.

Our approach is illustrated at a high level in Figure~\ref{fig:vid_net}.
Video \textit{interpolation} aims to predict a set of unseen intermediate frames 
from a pair of bordering reference frames.
Video \textit{extrapolation}, on the other hand, 
forecasts unobserved video frames based on those that have occurred in the past.
In our approach, an encoder $E$ learns how to produce a binary motion encoding, 
shown in the middle of the figure, 
with binarisation performed directly within the neural network. 
The resulting learned binary motion code is subsequently used 
to guide the extrapolation of P-frames conditioned on a past I-frame, 
or the interpolation of B-frames conditioned on bounding past and future I-frames.
Interpolation or extrapolation is performed by the decoder $D$ and the conditioning 
is indicated through the `$\textrm{Cond}$' block in the figure,
which extracts features from I-frames that have been compressed and decompressed 
independently by an existing image codec.
We therefore view the decoding of P-frames or B-frames in video compression 
as motion guided interpolation or extrapolation, 
where a low dimensional learned binary motion code helps direct prediction from I-frames.

\subsection{Related Work}
In work interested purely in prediction (without compressing), 
deep learning has been shown to produce high quality video frame 
interpolations~\cite{Jia2016,Liu2017,Niklaus2017,Niklaus2018,Jiang2018,Meyer2018,Bao2019,Liu2019} 
and extrapolations~\cite{Mathieu2016,Vondrick2016,Xue2016,Finn2016} for small time-steps.
Typically, unseen video frames are predicted 
solely based on the reference frame~\cite{Liu2017,Jiang2018,Meyer2018}.
For predicting unseen frames over longer time-spans 
(as would be the case if we were interested in video compression), 
additional information is required.

In video compression, we do not need to rely solely on the reference frame content: 
we can estimate the motion from the actual unseen video frames, 
compress these motion encodings, 
and then transmit this together with the compressed versions of the reference frames.
This extra motion information can enable video frame prediction 
over extended timespans~\cite{Wu2018}.
Based on this idea, deep learning has recently been applied to video compression, 
producing models capable of outperforming standard video codecs (H.264, H.265) 
at certain bit allocations~\cite{Rippel2018,Lu2019,Wu2018,Chen2018,Habibian2019,Cheng2019}.
These models combine state-of-the-art image compression, 
flow prediction and entropy coding networks to produce end-to-end 
optimisable video compression systems.
Despite their success, 
whole systems are evaluated as a single unit, 
making it difficult to discern to what extent each individual component outperforms 
its more conventional standard implementation.
In this work, we focus specifically on motion compression for video prediction---we 
consider P- and B-frame prediction in isolation, 
decoupled from all other compression components.

In~\cite{Wu2018}, video frames are hierarchically interpolated by warping input reference 
frame features with standard block-MVs, 
while discrete representations of motion are learned by encoding 
optical flow patterns in~\cite{Rippel2018,Lu2019}.
Optical flow vectors describe how pixels in a video frame should 
be moved over time to best estimate true object and camera motion~\cite{Horn1981}.
It is effective at modeling translational motion, 
but fails to capture more complex transformations such as rotation, 
warping, occlusion and changes in lighting~\cite{Richards2010}.
\cite{Rippel2018,Lu2019,Wu2018} address this by jointly compressing 
the residuals produced after flow compensation. 
In this paper, rather than using optical flow, 
we show that it is possible to learn compact encodings
that are representative of complex motion directly from a video sequence.
More specifically, we train an encoder network to produce learned binary motion codes which guide 
the prediction of P-frames and B-frames from I-frame context at the decoder.
Experiments show that the complex motion contained in our 
binary motion codes outperforms that of conventional optical flow.
The codes produced by our network could, therefore, 
provide an alternate means of motion conditioning for applications 
that are currently reliant on optical flow-based methods~\cite{Rippel2018,Lu2019,Wu2018,Ho2019}.

Different spatial and temporal locations in a 
video sequence are not necessarily equally complex.
In image compression, it has been shown that compression rates can be improved 
by varying bitrates such that less complex image regions are assigned fewer bits~\cite{Johnston2018,Minnen2018,Li2018,Balle2018,Lee2019}.
Varying the bitrate temporally in accordance to complexity of motion 
is equally important in video compression~\cite{ITU-T2003,ITU-T2018,Google2016}.
Most videos contain still segments interspersed with sequences depicting rapid motion.
A model with access to reference frame context at its decoder should learn to encode 
very little information for still video segments 
and allocate the bulk of its bits to intervals containing a high degree of motion.
In~\cite{Rippel2018,Han2018} recurrence is used to sequentially maintain state 
information across time such that previously 
decoded information need not be re-encoded.
We, on the other hand, 
extend the 2D content-weighted image compression technique in~\cite{Li2018} 
to three dimensions and present parallelised P-frame and B-frame 
video compression models that learn to vary bitrate both spatially and temporally.
Experiments show how our approach to 3D dynamic bit assignment 
substantially reduces the bitrate of a motion encoding model 
without adversely affecting its reconstruction quality.

\subsection{Contributions}
We proceed as follows. 
First we give a detailed description of our 
P-frame and B-frame compression architectures.
We then formulate our approach to 3D content-aware bit weighting and 
demonstrate its applicability to bitrate optimisation.
Finally, compression efficiency is evaluated in terms of various 
video quality metrics (PSNR, SSIM, VMAF and EPE). 
We demonstrate that our models' P-frame and B-frame predictions outperform those of the
block-motion prediction algorithms employed by standard video codecs such as H.264 and H.265.
Additional experiments are carried out to determine the impact of 
an optical flow-based loss term 
and if multi-scale convolutions result in richer motion sampling.
We find that including multi-scale convolutions in our encoder architecture
slightly improves the quality of our model's video frame predictions.
On the other hand, limiting our model to learning 
pixel-wise translational motion with a flow loss term 
worsens its prediction quality. 
This indicates that we are able to learn more 
representative motion than conventional optical flow.
A full implementation of the code developed as part of this work is made available online: \href{https://github.com/adnortje/deepvideo}{DeepVideo}.\footnote{
\url{https://github.com/adnortje/deepvideo}}


%% file: my_content/arch.tex
%

\section{Video Frame Prediction Architecture}

\subsection{Architectural Overview}
\label{sec:arch_ovr}
Figure~\ref{fig:vid_net} illustrates our approach 
to P-frame and B-frame prediction.
The neural network encoder $E$ compresses and binarises the motion 
occurring in a Group Of Pictures (GOP): 
a video segment containing designated reference (I) 
and referencing (P or B) frames.
Binarisation via thresholding is non-differentiable. 
To perform this operation directly within a neural network, 
we therefore resort to the stochastic binarisation function presented in~\cite{Toderici2015}: 
during training, each encoder output, which lies within $(-1,1)$,
is made to take one of two distinct values in the set $\{-1, 1\}$ 
by adding uniform quantisation noise.
This allows for a straight through estimate~\cite{Raiko2014} of gradients, 
i.e. gradients flow through the binarisation layer unchanged.
At the decoder $D$, I-frame features extracted by the conditioning network 
`$\textrm{Cond}$' are transformed based on the information held in 
the binarised motion encoding to predict the P or B referencing frames. 
Note that `$\textrm{Cond}$' is not responsible for I-frame compression: 
this is done by an existing image codec.

%
\begin{figure}[h]
	\center{
		\includegraphics[width=.85\textwidth]{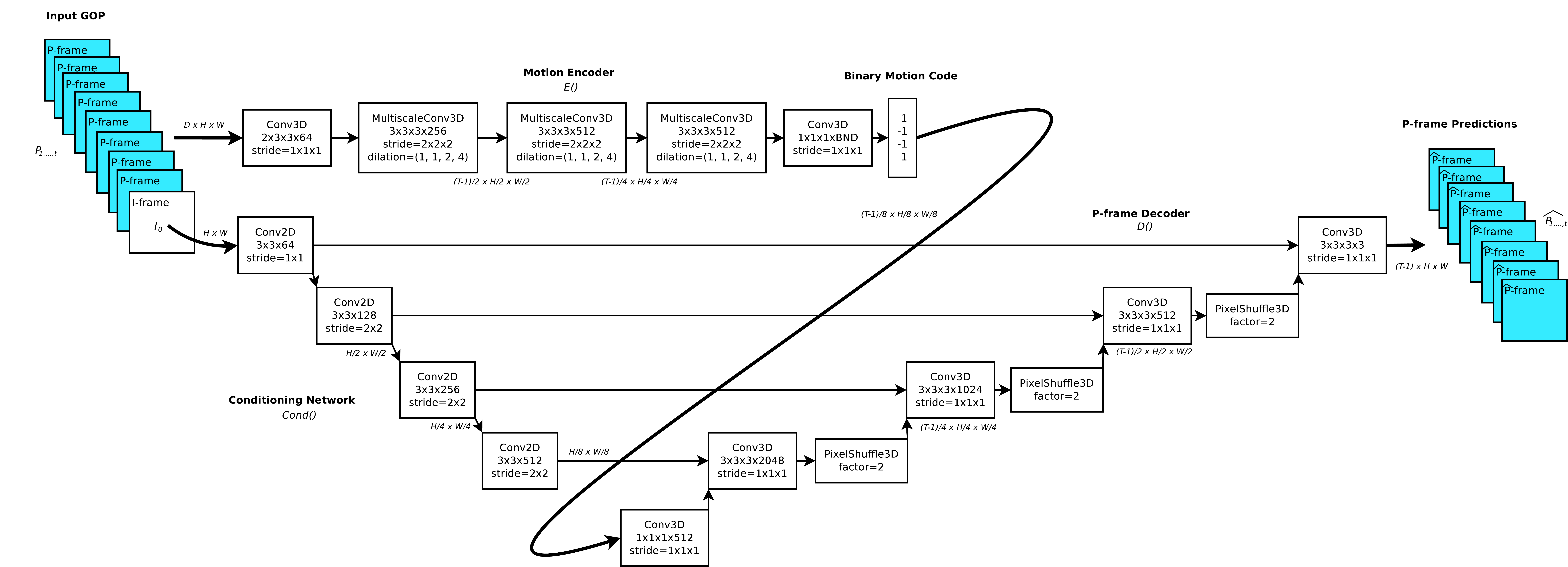}
	}
	\caption{
		The P-frame prediction network (P-FrameNet) 
		used to	extrapolate video frames from a past I-frame.
	}
\label{fig:pf_auto}
\end{figure}
%

%
\begin{figure}[h]
	\center{
		\includegraphics[width=.85\textwidth]{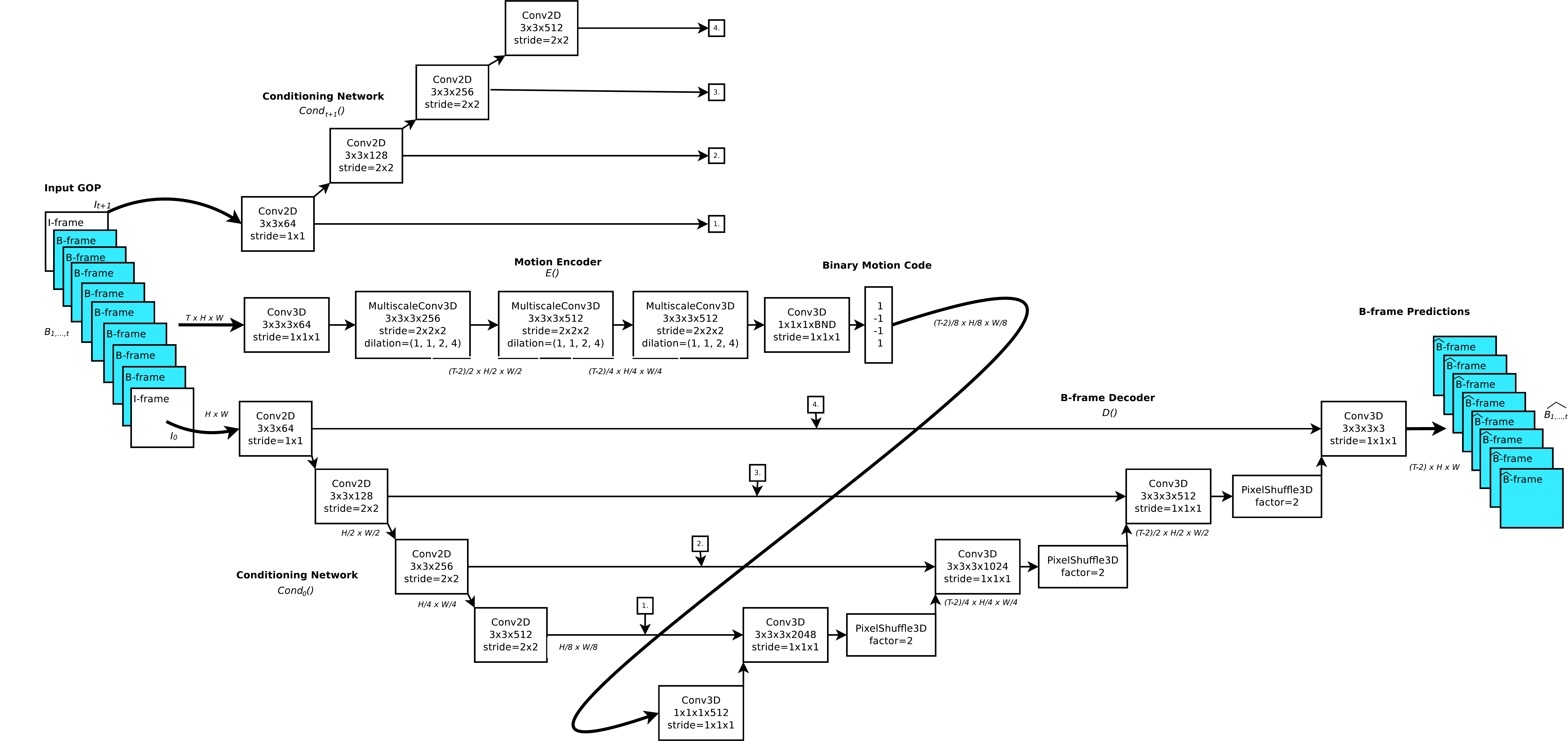}
	}
	\caption{
		The B-frame prediction network (B-FrameNet)
		used to bi-directionally interpolate video frames from 
		bounding past and future I-frames.
	}
\label{fig:bf_auto}
\end{figure}

\subsection{P-frame and B-frame Prediction Networks}
\label{sec:pf_and_bf_nets}
Figures~\ref{fig:pf_auto} and~\ref{fig:bf_auto} illustrate our 
P-frame and B-frame prediction networks in greater detail.
Equations~\eqref{eq:pf_auto} and~\eqref{eq:bf_auto} summarise the P-frame 
and B-frame prediction processes depicted in Figures~\ref{fig:pf_auto} 
and~\ref{fig:bf_auto}, respectively.
%
\begin{equation}
	\widehat{P_{1,\ldots,t}} 
	= D\left(E(I_0, P_{1,\ldots,t}), \textrm{Cond}(I_0) \right)
	\label{eq:pf_auto}
\end{equation}
%
%
\begin{equation}
	\widehat{B_{1,\ldots,t}} 
	= D\left(E(I_0, B_{1,\ldots,t}, I_{t+1}), 
	\textrm{Cond}_0(I_{0}), \textrm{Cond}_t(I_{t+1})\right)
	\label{eq:bf_auto}
\end{equation}
The decoder $D(\cdot)$ uses context derived from reference frames 
$I$ by the conditioning network $\textrm{Cond}(\cdot)$ 
to predict a sequence of $t-1$ frames, 
$P_{1,\ldots,t}$ or $B_{1,\ldots,t}$.
The prediction process is supervised by a binarised motion encoding, 
$E(\cdot)$, of the original GOP sequence.
Because the encoder always compresses the input GOP's width, 
height and time axes by a factor of 8, 
P-FrameNet and B-FrameNet's bitrate is determined by 
the number of output channels we set in the final encoder layer, $C_{\textrm{bnd}}$.
We denote predicted video-frames as either $P$ or $B$,
depending on whether the decoder performs motion guided 
extrapolation or interpolation.
The decoder performs extrapolation, Figure~\ref{fig:pf_auto}, 
when conditioned on a single I-frame, $I_0$,
and interpolation, Figure~\ref{fig:bf_auto}, 
when conditioned on a pair of bounding I-frames, $I_0$ and $I_{t+1}$.
During training we use a $L_2$ reconstruction loss:
%
\begin{equation}
	L_R = \abss{B - \hat{B}}^{2} \quad 
	\text{or} \quad \abss{P - \hat{P}}^{2},
	\label{eq:r_loss}
\end{equation}
Throughout training we give $D(\cdot)$ access to the original I-frame content, 
but at test time $I_0$ and $I_{t+1}$ are encoded 
and decoded independently by an existing image codec.

Motion in video often occurs at different scales.
To account for this, we implement 3D multi-scale convolutional 
layers~\cite{Yu2016,Baig2017} in our encoder network as a lightweight 
substitute for deep pyramidal decomposition~\cite{Mathieu2016,Rippel2017}.
Each multi-scale convolution combines filters with different dilation factors 
for more diverse motion sampling across a range of spatial and temporal scales, 
and can be seen as a type of learned scale invariant feature transform (SIFT)~\cite{Lowe2004}.
In Section~\ref{sec:ms_conv_exp} we demonstrate how the inclusion of multi-scale convolutions consistently improves video frame prediction quality.
As shown in Figures~\ref{fig:pf_auto} and~\ref{fig:bf_auto}, 
manifold layers from the conditioning network `{\textrm{Cond}}'
are joined to the decoder {$D$} in a fashion reminiscent 
of the U-Net~\cite{Ronneberger2015} architecture, 
prevalent in previous video interpolation work~\cite{Liu2017,Jiang2018}.
I-frame conditioning at the decoder enables P-FrameNet and B-FrameNet 
to learn motion compensation (how to transform I-frame content) 
instead of just compressing input P- and B-frames directly. 
The experiment in Section~\ref{sec:cond_exp} shows that the binary motion codes learnt 
through I-frame conditioning are more easily compressible than raw video frames.
Upscaling at the decoder is accomplished via pixel-shuffling~\cite{Shi2016}, 
an efficient alternative to transposed convolutions.
Scene changes and motion complexity often dictate GOP length 
selection in standard codecs~\cite{Richards2010}. 
Our designs are, therefore, 
fully convolutional to ensure that they are able
to accommodate a diverse range of input frame-sizes and dynamic GOP lengths.

\subsection{3D Dynamic Bit Assignment}
\label{sec:dba_arch}
In order to vary the bitrate of our binary motion codes, 
we leverage~\cite{Li2018}'s approach to content-weighted image compression 
and \textit{learn} to vary bitrate across an extra dimension: time.
Video regions that are smooth and predominantly stationary are easier 
to compress than those containing rich texture and rapid motion.
An ideal motion compression model should, therefore,
actively adapt its bitrate according to fluctuations in video complexity
by assigning fewer bits to simplistic video regions and vice versa.
As it stands, our encoder architecture allocates a fixed number of bits to 
each spatio-temporal location in its code-space, 
specified by $C_{\textrm{bnd}}$, the number of channels in its binarisation layer.
Based on~\cite{Li2018}, we learn a 3D bit-distribution-map, $B_{\textrm{map}}$, 
that determines how many bit channels are allocated 
to our binary motion encoding per point in space-time.

\begin{figure}[!h]
        \center{
                \includegraphics[width=.99\textwidth]{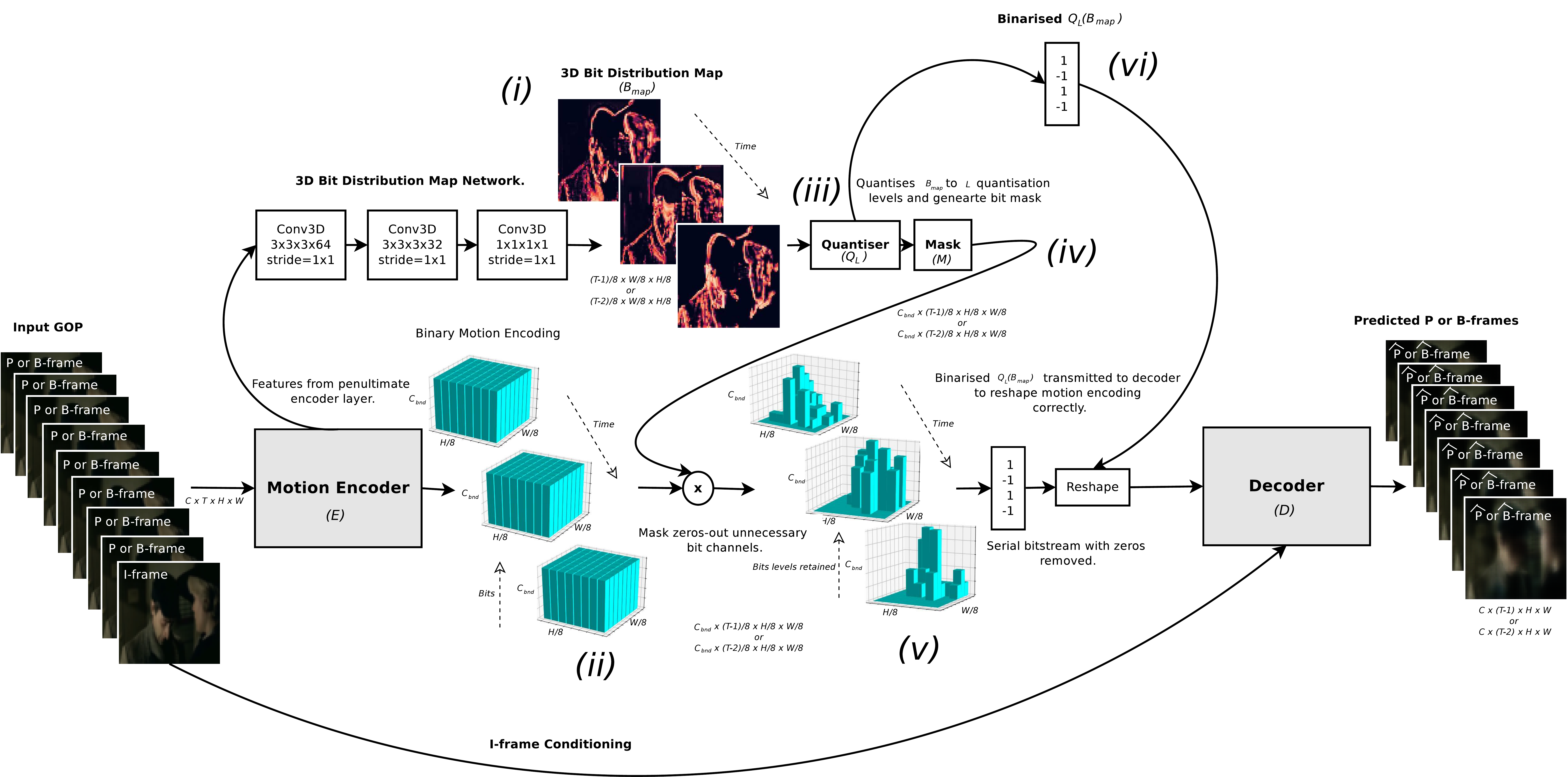}
        }
        \caption{
                3D dynamic bit assignment incorporated into a video frame 
                prediction model to vary its bit allocations across space-time.
                In this figure the motion encoding bit-space is represented in its 
                true multi-dimensional form by blue blocks.
                $B_{\textrm{map}}$, indicated by \textit{(i)} in the figure, 
                is used to generate a mask $M$ at \textit{(iv)} that crops out 
                unnecessary bits at \textit{(v)}.
        }
\label{fig:dba_net}
\end{figure}

Figure~\ref{fig:dba_net} illustrates the key stages in 
our approach to 3D dynamic bit assignment.
First we learn $B_{\textrm{map}}$, shown at \textit{(i)} in the figure, 
from the input GOP by passing features extracted 
by the penultimate encoder layer through a 3D convolutional network.
$B_{\textrm{map}}$ is a single-channel feature map whose values fall in the range $(0, 1)$ 
and whose spatial and temporal size is the same as the binary motion code produced by the encoder, 
represented by the blue cubes at \textit{(ii)}.
While in the figures thus far we have indicated the serialised motion 
encoding with a box of $-1$ and $1$s, 
the cubes at \textit{(ii)} in this figure indicate the individual learned MVs
for each video frame over its width and height 
(these are serialised later, as explained below).
The lighter regions in $B_{\textrm{map}}$ are higher valued 
and identify video regions that should be allocated more bits (channels).
Following~\cite{Li2018}, we portion the available $C_{\textrm{bnd}}$ bits produced 
for each video frame by the encoder into $L$ groups each containing $\frac{C_{\textrm{bnd}}}{L}$ bits.
With $\floor{\cdot}$ denoting the mathematical floor operator, 
each element, $b_{t,h,w}$, in $B_{\textrm{map}}$ is quantised to one of $L$ integer levels,
%
\begin{equation}
	Q_{L}(b_{t,h,w}) = \floor{Lb_{t,h,w}} 
	\label{eq:quant}
\end{equation} 
to decide how many bit levels need to be retained per point in space-time.
To avoid allocating non-integer bit numbers 
we require that $C_{\textrm{bnd}}$ be cleanly divisible by $L$ and $L \leq C_{\textrm{bnd}}$.
Guided by $Q_{L}(B_{\textrm{map}})$ at \textit{(iii)}, we populate a mask $M$, 
shown at \textit{(iv)}, 
that zeros-out unnecessary bit channels produced by the encoder at \textit{(ii)}:
%
\begin{equation}
	m_{c,t,h,w} = 
		\begin{cases}
		1, & \text{if $c\leq\frac{C_{\textrm{bnd}}}{L}Q_L(b_{t,h,w})$} \\
		0, & \text{otherwise}
		\end{cases}
	\label{eq:bit_mask}
\end{equation}
The cubes at \textit{(v)} shows how masked bits (zeros) are cropped-out 
prior to the transmission of the serialised motion bitstream. 
Zeros are reinstated at the decoder by zero-padding 
each channel to $C_{\textrm{bnd}}$ (the maximum bit-length).
After multiplication by $M$ and zero-cropping,
the number of bits transmitted per point in space-time is reduced from
$C_{\textrm{bnd}}$ to $\frac{C_{\textrm{bnd}}}{L}Q_L(b_{t,h,w})$.
In order for the decoder to reshape the serial bitstream correctly, 
a binarised version of $Q_L(B_{\textrm{map}})$ is sent separately as 
additional overhead at \textit{(vi)}.
The integer values in $Q_L(B_{\textrm{map}})$ 
are binarised using base-2 expansion~\cite{Lathi2018} for transmission.
 
Realising that our serial bit-count is proportional to the summation over $B_{\textrm{map}}$,
we can use an additional loss term,
%
\begin{equation}
	L_{B} = \sum_{t,h,w}b_{t,h,w}
	\label{eq:rate_loss}
\end{equation}
to drive down our model's bitrate during training~\cite{Li2018}.
$L_{B}$ penalises bitrates above zero.
This prevents assigning bits to stationary video regions 
that can be deduced from I-frame context alone.

Both the mask formation, \textit{(iv)}, and quantisation functions, \textit{(iii)},
in equations~\eqref{eq:quant} and~\eqref{eq:bit_mask} are non-differentiable.
Luckily, using straight through estimation~\cite{Li2018,Raiko2014} again, 
the gradient of $M$ with respect to $b_{t,h,w}$ can be approximated by,
%
\begin{equation}
	\frac{\partial m_{c,t,h,w}}{\partial b_{t,h,w}} = 
		\begin{cases}
		L, & \text{if $Lb_{t,h,w} - 1 \leq \ceil{\frac{cL}{C_{\textrm{bnd}}}} \leq Lb_{t,h,w}+2$} \\
		0, & \text{otherwise}
		\end{cases}
	\label{eq:bit_mask_deriv}
\end{equation}
We show the benefit of this 3D dynamic bit assignment 
approach experimentally in Section~\ref{sec:3d_dba_exp}.

%

\subsection{Optical Flow Loss}
\label{sec:flow_loss_arch}
\begin{figure*}[!h]
	\center{
		\includegraphics[width=.99\textwidth]{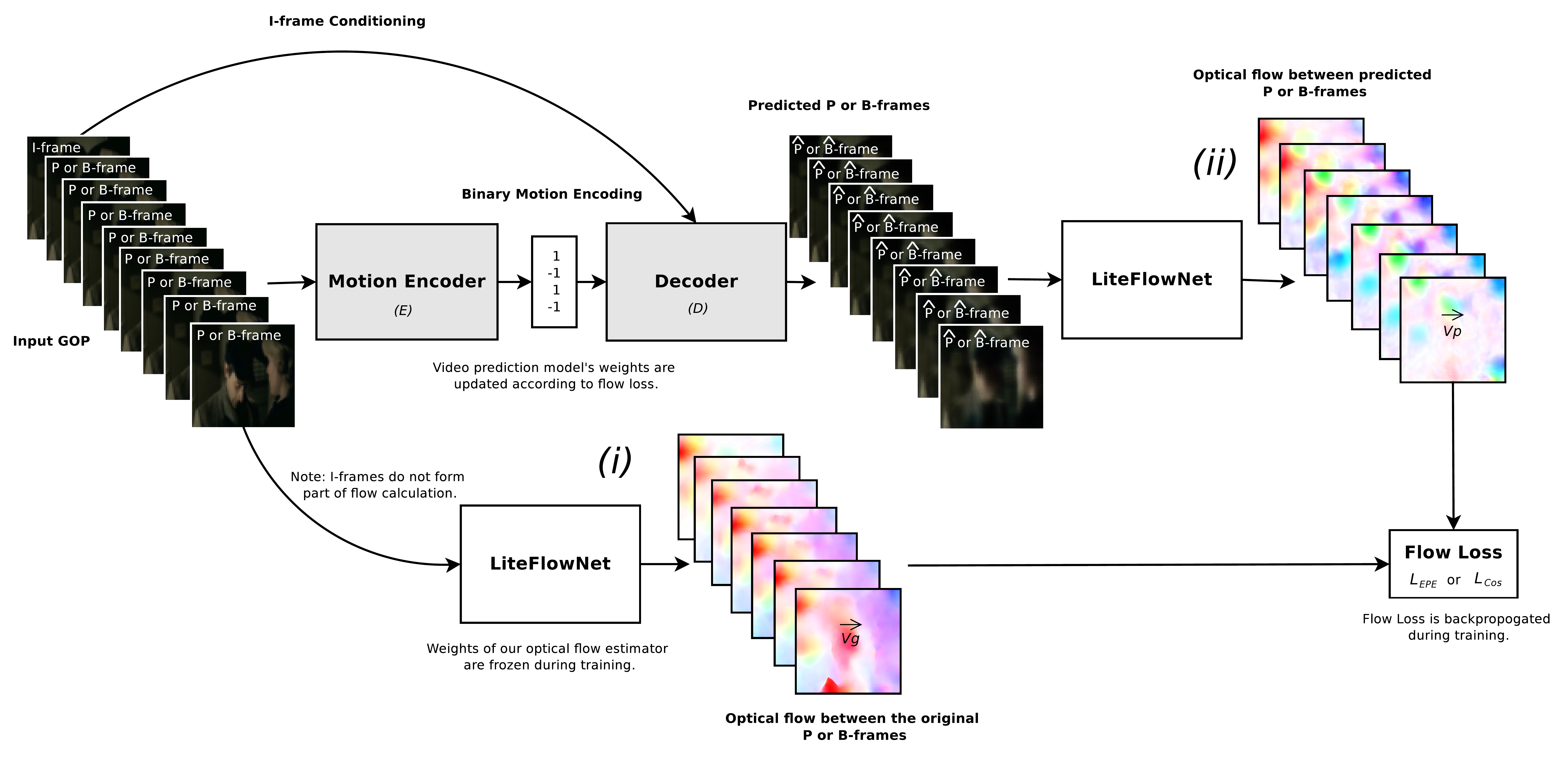}
	}
	\caption{
		Setup used to train a video frame prediction network 
		with an optical flow based loss term. 
		We use LiteFlowNet~\cite{Hui2018} to calculate and compare the optical 
		flow of the input and predicted video frames. 
		LiteFlowNet's weights are fixed when optimising our video frame prediction models.
	}
\label{fig:flow_setup}
\end{figure*}

We deploy the setup shown in Figure~\ref{fig:flow_setup} to determine if 
including an explicit additional optical flow based loss term 
leads to improved motion compression.
The optical flow between two video frames is defined by a 2D vector field 
that relates the movement of pixels from 
the one frame to the other~\cite{Richards2010}.
We denote the dense (per-pixel) optical flow
for each consecutive pair of frames in the input GOP as 
$\vec{V_g}$: the ground truth flow, 
indicated at \textit{(i)} in the figure.
As shown at \textit{(ii)}, 
$\vec{V_p}$ represents the flow vectors derived from
the frames predicted by our motion compression network.
A host of techniques can be used to calculate $\vec{V_g}$ 
and $\vec{V_p}$, including differential~\cite{Farneback2003}, 
phase~\cite{Gautama2002} and energy~\cite{Horn1981} based methods, 
or more recent deep learning approaches~\cite{Fischer2015,Ilg2016,Zweig2016,Ranjan2017,Hui2018}.
In this work, 
we use LiteFlowNet~\cite{Hui2018}, 
a state-of-the-art deep flow estimation model.
LiteFlowNet's weights are pre-trained on 
the MPI Sintel dataset~\cite{Butler2012} 
and frozen when training our video compression models.
We experiment with the optical flow losses defined by the Euclidean distance between 
the ground truth and predicted flow vectors called the end-point-error (EPE),
%
\begin{equation}
	L_{\textrm{EPE}} = \sqrt{\abss{\vec{V_g} - \vec{V_p}}^2},
	\label{eq:epe_loss}
\end{equation}
and cosine similarity,
%
\begin{equation}
	L_{\textrm{cosine}} 
	= 1 - \frac{\vec{V_g} \cdot \vec{V_p}}
	{\|\vec{V_g}\| \ \|\vec{V_p}\|}.
	\label{eq:cos_loss}
\end{equation}
$L_{\textrm{cosine}}$ differs from $L_{\textrm{EPE}}$ in that it 
only penalises directional deviations between 
the ground-truth and predicted flow vectors
as disregarding differences in magnitude may provide beneficial regularisation.
We normalise the $x$ and $y$ components of 
the flow vectors in $\vec{V_g}$ and $\vec{V_p}$ 
by the width and height of the input video frames to avoid directional biasing.
We investigate the consequences of adding these optical 
flow loss terms in Section~\ref{sec:flow_loss_exp}.

%
%


%% file: my_content/train.tex

\section{Experimental Setup}

\subsection{Data and Training Procedure}
The P-frame and B-frame prediction networks 
in Figures~\ref{fig:pf_auto} and~\ref{fig:bf_auto}
are trained on the Hollywood dataset~\cite{Hollywood2009}.
This dataset contains 475 AVI movie clips 
from a wide range of classic films.
The average clip length in our training corpus is around 5 seconds.
Prior to training we transcode each clip with 
the H.264~\cite{ITU-T2003} codec to ensure NVIDIA Video Loader (NVVL) 
data loader compatibility~\cite{NVIDIA2018,NVIDIAG2018}.
We split this dataset into training and validation sets 
containing 435 and 40 clips, respectively.
To avoid learning compression artefacts introduced by H.264, 
we train our models on resized $64\times64$ pixel video frames.
During training we randomly crop a GOP from each clip, 
so although our dataset only contains 435 videos, 
our models are exposed to substantially more data.
The GOPs used for validation are cropped from the start of each video 
in the validation set to ensure that the validation losses 
used for early stopping are directly comparable across epochs.
GOP length is set to 18 for B-FrameNet and 17 for P-FrameNet.
In both cases our models are trained to predict 16 video frames 
using the reconstruction loss in equation~\eqref{eq:r_loss}.
Input video clips are grouped into batches of 3 
and their pixel values are normalised to fall in the range $(-1,1)$.
The training process spans 150 epochs 
and utilises Adam optimisation~\cite{Kingma2014}.
The initial  learning rate is set to 0.0001 
and decayed by a factor of 2 at epochs 30, 100 and 140.
We train models with a range of bottleneck-depths $C_{\textrm{bnd}}$ to 
gauge performance across a range of bitrates.

\subsubsection{Bitrate Optimisation}
\label{sec:dba_train}
Pre-trained P-FrameNet and B-FrameNet models are trained to perform 3D 
dynamic bit assignment (Section~\ref{sec:dba_arch}) by undergoing another 150 epochs of training 
with the optimisation function defined by
%
\begin{equation}
	L_{RB} = L_{R} + \lambda L_{B}
\label{eq:rate_rec_loss}
\end{equation} 
$L_{RB}$ combines the reconstruction loss 
$L_R$ in equation~\eqref{eq:r_loss} with the loss $L_{B}$ 
in equation~\eqref{eq:rate_loss} that encourages low bitrates.

As done in~\cite{Li2018}, 
we introduce the hyperparameter $\lambda$ to control 
the trade-off between reconstruction quality and compression rate.
Setting $\lambda=0$, zeros-out $L_B$, 
which prevents a model from learning to vary its bitrate.

\subsubsection{Flow Loss}
\label{sec:flow_train}
When training our models to minimise the difference between 
the optical flow of the input 
and that of the predicted frames 
(Section~\ref{sec:flow_loss_arch}) we use the loss defined by,
%
\begin{equation}
	L _{RF}= L_R + \alpha L_F
\label{eq:flow_rec_loss}
\end{equation}
where $L_F$ is one of the the optical flow losses 
in equations~\eqref{eq:epe_loss} or~\eqref{eq:cos_loss} 
and $L_R$ is the distortion loss in equation~\eqref{eq:r_loss}.
$\alpha$ is a weighting term used to normalise $L_F$ by the 
total number of flow vectors during training.

\subsection{Evaluation} 
To quantify the quality of our predicted video frames 
we use three objective evaluation metrics: 
Peak Signal to Noise Ratio (PSNR),
Structural SIMilarity index (SSIM)~\cite{Wang2004},
and Video Multi-method Assessment Fusion (VMAF)~\cite{Netflix2018}.
PSNR and SSIM measure the degree to which an image reconstruction 
corresponds to the original image, so we calculate PSNR 
and SSIM for video by averaging scores across the predicted video frames.
SSIM falls within $[-1,1]$ while PSNR can be any real value 
and is expressed in dB.
Higher scores signify greater prediction quality. 

We also assess our video compression with the VMAF
framework developed and deployed by Netflix~\cite{Netflix2018}.
VMAF is a machine-learning based video quality metric 
trained to combine the results of various perceptual 
models such that its scores are more 
closely aligned with the human visual system
than a stand-alone objective algorithm~\cite{Netflix2018}. 
VMAF scores fall in the range $[0, 100]$, 
with a higher score again 
indicating greater reconstruction quality~\cite{NetflixG2018}.

To understand and probe different aspects of our approach, 
the experiments in Section~\ref{sec:exp} 
are carried out on videos from our validation set: 
the 40 videos from the Hollywood dataset~\cite{Hollywood2009}.
In Section~\ref{sec:results} final versions of P-FrameNet and B-FrameNet 
are pitted against the block-motion algorithms 
used in standard video codecs.
This evaluation is carried out on 235 raw 
(\sq{\texttt{.yuv}}) video clips sampled 
from the Video Trace Library (VTL)~\cite{VTL2000}.\footnote{
	Additionally, we provide YouTube links to videos compressed 
	by P-FrameNet/B-FrameNet that have been taken from the wild.
}
Each clip is partitioned into 17-frame or 18-frame sequences depending on whether 
we are predicting P-frames or B-frames, 
such that our models always predict 16-frames.
VMAF, SSIM, PSNR and EPE scores are then calculated 
and averaged across the reconstructed video-frames.
We denote EPE as EPE (FlowNet) or EPE (Farneback) depending on whether 
we calculate optical flow using LiteFlowNet~\cite{Hui2018} 
or Farneback's polynomial method~\cite{Farneback2003}; 
in contrast to the other metrics, lower EPE is better.


%% file: my_content/exp.tex
\section{Results: Ablation Experiments and Analysis}
\label{sec:exp}

In order to probe and better understand our approach through ablation studies 
and developmental experiments, 
we guide our analysis using the following questions.

\subsection{P-frame vs.\ B-frame Decoder Conditioning?}
\label{sec:cond_exp}
We explore the benefits of conditioning our video decoder on learned 
features extracted from reference I-frames (Section~\ref{sec:pf_and_bf_nets}).
Table~\ref{tab:cond_exp} compares a video autoencoder 
(P-FrameNet without I-frame conditioning) to P-FrameNet 
(single reference frame conditioning) 
and B-FrameNet (dual reference frame conditioning) 
in Figures~\ref{fig:pf_auto} and~\ref{fig:bf_auto}, respectively. 
In Table~\ref{tab:cond_exp} conditioning is shown to consistently 
improve the quality of the predicted frames as it allows the encoder 
to focus primarily on motion extraction---we 
learn to transform available pixel-content rather than compressing it directly.

Table~\ref{tab:cond_exp} reaffirms that motion transforms 
are easier to compress than raw video content~\cite{Richards2010}. 
B-frame conditioning is shown to outperform its P-frame counterpart, 
as context from bounding reference frames allows it to learn both
forward and reverse motion transformations.
Unlike standard video codecs, 
B-FrameNet is able to predict B-frames in parallel without the extra overhead of 
having to transmit the order in which frames are to be decoded~\cite{ITU-T2018}.
%
\begin{table}[!h]
\caption{Quality scores for various decoder conditioning schemes.}
	\begin{center}
		\begin{tabularx}{.7\linewidth}{lCCC}
			\toprule
				Conditioning & PSNR & SSIM & VMAF \\
			\midrule
				None	& $23.71$ & $0.64$ & $41.21$ \\
				P-frame	& $28.25$ & $0.80$ & $62.31$ \\
				B-frame	& $\mathbf{29.83}$ & $\mathbf{0.84}$ & $\mathbf{70.45}$ \\
			\bottomrule
		\end{tabularx}
	\end{center}
\label{tab:cond_exp}
\end{table}

\subsection{Do Multiscale Convolutions Learn More Representative Motion?}
\label{sec:ms_conv_exp}
We experiment with incorporating the multi-scale 
convolutions~\cite{Yu2016} discussed in Section~\ref{sec:pf_and_bf_nets} 
in our motion encoder architecture.
This provides us with a lightweight means of sampling motion 
at a variety of spatial and temporal scales.
Table~\ref{tab:multiscale_exp} compares two implementations of P-FrameNet, 
one with multi-scale convolutions and the other with normal convolutions in its motion encoder.
Using multi-scale convolutions leads to modest but consistent improvements in the quality 
of the predicted frames, as indicated by higher PSNR, SSIM and VMAF scores. 
It also allows P-FrameNet to perform more representative 
motion encoding (lower EPE).
Multi-scale convolutional layers are, therefore, used in 
our motion encoders throughout the rest of this work.

\begin{table}[!h]
\caption{Multis-scale vs.\ standard convolutional implementations of P-FrameNet.}
	\begin{center}
		\begin{tabularx}{.95\linewidth}{@{}lCCCCC@{}}
			\toprule
				\multirow{2}{*}{Convolution} & 
				\multirow{2}{*}{PSNR} & 
				\multirow{2}{*}{SSIM} & 
				\multirow{2}{*}{VMAF} & 
				\multicolumn{2}{c}{EPE} \\
				\cmidrule{5-6}
				&&&& \small{FlowNet} & \small{Farneback} \\
			\midrule
				Standard & $28.25$ & $0.80$ & $62.31$ & $0.477$ & $9.28 \cdot 10^{-7}$ \\
				Multiscale & $\mathbf{28.48}$ & $\mathbf{0.81}$ & $\mathbf{63.90}$ & $\mathbf{0.471}$ & $\mathbf{9.24 \cdot 10^{-7}}$ \\ 
			\bottomrule
		\end{tabularx}
	\end{center}
\label{tab:multiscale_exp}
\end{table}

\subsection{Is an Optical Flow Based Loss Beneficial?}
\label{sec:flow_loss_exp}
We experiment with the EPE and cosine similarity flow losses in 
equations~\eqref{eq:epe_loss} and~\eqref{eq:cos_loss} in Section~\ref{sec:flow_loss_arch} 
to discover if an optical flow based penalty helps B-FrameNet to learn improved motion.
As stated in equation~\eqref{eq:flow_rec_loss} in Section~\ref{sec:flow_train}, 
we add our chosen flow loss, $L_F$, to the reconstruction loss, $L_R$, 
during training. 
The hyperparameter $\alpha$ in equation~\eqref{eq:flow_rec_loss} is used to weight $L_F$ 
such that the mean loss per flow vector is added to $L_R$. 

Table~\ref{tab:flow_exp} reveals that an additional optical flow loss term
worsens the quality of B-FrameNet's reconstructions (lower PSNR, SSIM and VMAF scores). 
The added loss term does, however, 
cause the optical flow of the predicted frames to 
match that of the input more closely (lower EPE).
At first glance, 
this result seems contradictory. 
How can learning better motion lead to a depreciation in quality?
Realising that optical flow is essentially only 2D pixel shuffling, 
the results in Table~\ref{tab:flow_exp} imply that B-FrameNet is able to learn 
more advanced motion transforms (rotation, warping, occlusion and colour shift) 
when it is not limited to translational motion by a flow loss term.
Informal experiments showed that increasing the weight $\alpha$ of the flow loss term 
in equation~\eqref{eq:flow_rec_loss} improved the EPE but further degraded 
the quality of the predicted video frames.
%
\begin{table}[!h]
\caption{
	Influence of an additional EPE (equation~\eqref{eq:epe_loss}) 
	or cosine (equation~\eqref{eq:cos_loss}) 
	optical flow loss term on B-FrameNet's performance.
}
	\begin{center}
		\begin{tabularx}{\linewidth}{lCCCCC}
			\toprule
				\multirow{2}{*}{Flow Loss} & 
				\multirow{2}{*}{PSNR} & 
				\multirow{2}{*}{SSIM} & 
				\multirow{2}{*}{VMAF} & 
				\multicolumn{2}{c}{EPE} \\
				\cmidrule{5-6}
				&&&& \small{FlowNet} & \small{Farneback} \\
			\midrule
			None & $\mathbf{29.81}$ & $\mathbf{0.842}$ & $\mathbf{71.03}$ & $0.477$ & $9.21 \cdot 10^{-7}$ \\
			EPE & $29.59$ & $0.836$ & $70.19$ & $\mathbf{0.461}$ & $\mathbf{9.17 \cdot 10^{-7}}$ \\
			Cosine & $24.04$ & $0.652$ & $59.44$ & $0.514$ & $9.69 \cdot 10^{-7}$ \\
			\bottomrule
		\end{tabularx}
	\end{center}
\label{tab:flow_exp}
\end{table}

\subsection{What are Optimal Parameters for Learning 3D Dynamic Bit Assignment?}
\begin{figure}[!h]
	\center{
		\includegraphics[width=.95\textwidth]{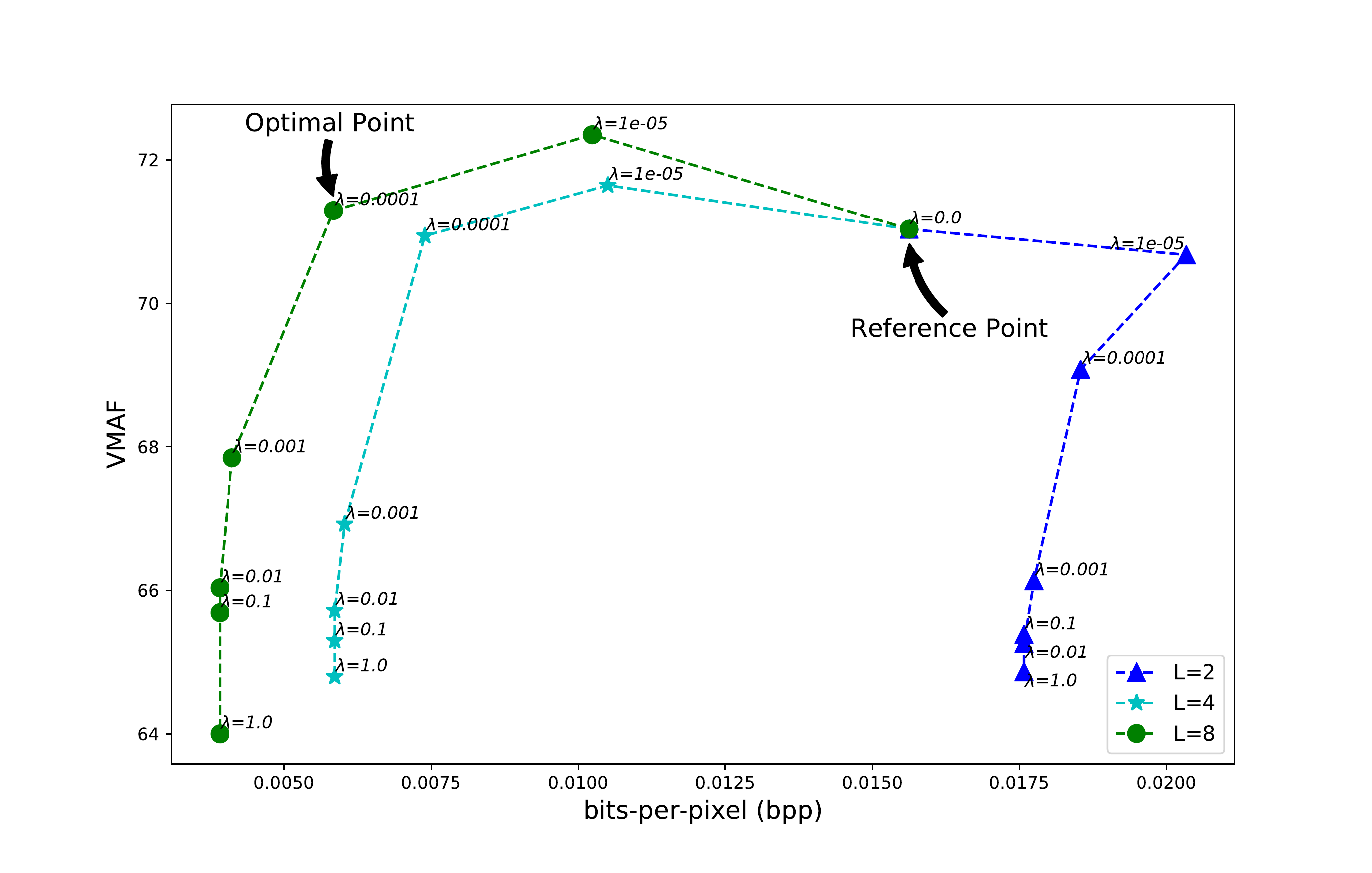}
	}
	\caption{
		VMAF scores and learned bitrates produced on development data
	 	by B-FrameNet trained with different parameterisations of $L_{RB}$.
	}
\label{fig:dbaL_exp}
\end{figure}

With this experiment we try to find an optimal parameterisation of 
the 3D dynamic bit assignment loss $L_{RB}$ in Section~\ref{sec:dba_train}.
We tune B-FrameNet ($C_{\textrm{bnd}}=8$) with different quantisation levels $L$ 
in equation~\eqref{eq:quant} and different values of $\lambda$  
for the weight of the binary penalisation 
term $L_B$ in the overall loss $L_{RB}$ in equation~\eqref{eq:rate_rec_loss}.
The learned bitrates and corresponding VMAF scores 
are shown in Figure~\ref{fig:dbaL_exp}.

Using $\lambda=0$ (no 3D dynamic bit assignment) as a point of reference, 
we find that setting $L$ equal to its maximum possible value, 
in this case the channel depth $8$, 
yields the best compression efficiency.
$L$ controls the number of bit levels 
the model can choose from per point in space-time. 
Higher settings of $L$ gives B-FrameNet more options in deciding 
how many bits to allocate to different video regions, 
bringing about a lower bitrate on average.
Setting $L$ too low, $L=2$, 
worsens compression as the bit-savings are not substantial enough to 
outweigh the cost of sending the quantised bit distribution map 
at \textit{(vi)} in Figure~\ref{fig:dba_net}.
We see that for $L=8$ and $L=4$, 
setting $\lambda=0.0001$ is optimal,
as it results in the greatest bit reduction without decreasing 
the quality of the predicted frames below that of 
the reference point, $\lambda=0$.
Based on these results we set $\lambda=0.0001$ 
and $L$ equal to our encoder's bottleneck-depth $C_{\textrm{bnd}}$ 
whenever optimising a system for 3D dynamic bit assignment.

%

\subsection{Does 3D Dynamic Bit Assignment Aid Motion Compression?}
\label{sec:3d_dba_exp}
\begin{figure}[!h]
	\centering
	\subfigure[PSNR]{
		\includegraphics[width=.4\textwidth]{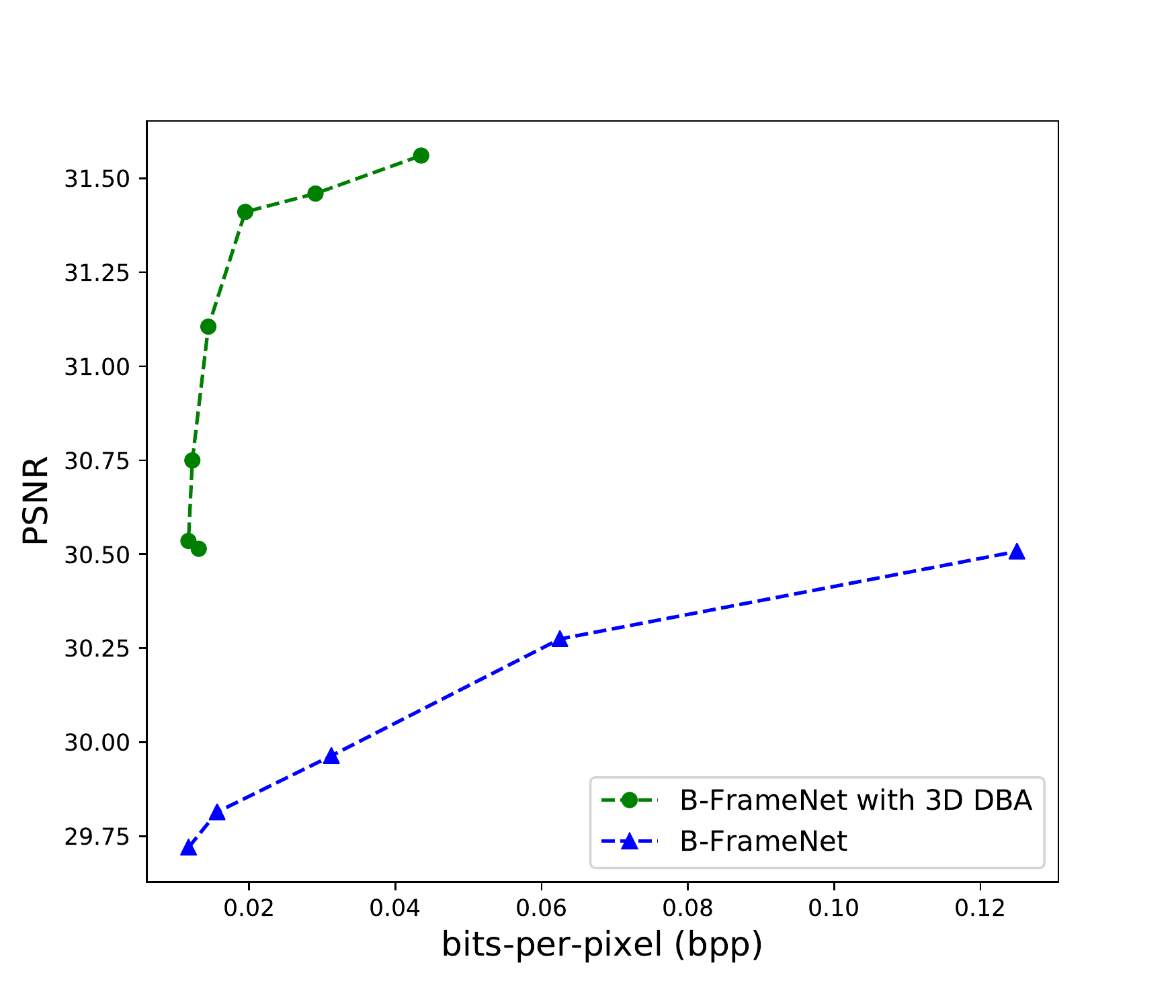}
	}
	\subfigure[SSIM]{
		\includegraphics[width=.4\textwidth]{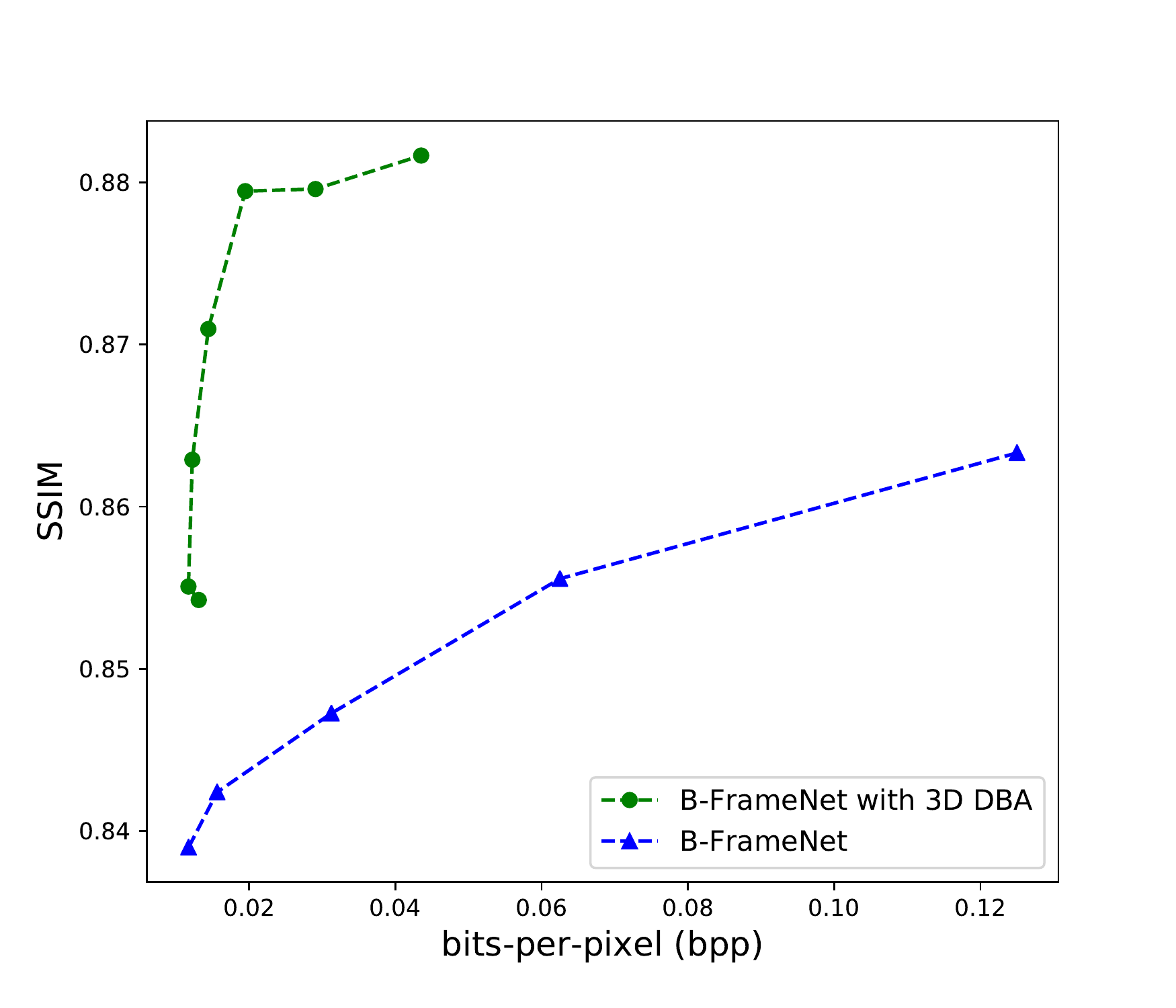}
	}
	\subfigure[VMAF]{
		\includegraphics[width=.4\textwidth]{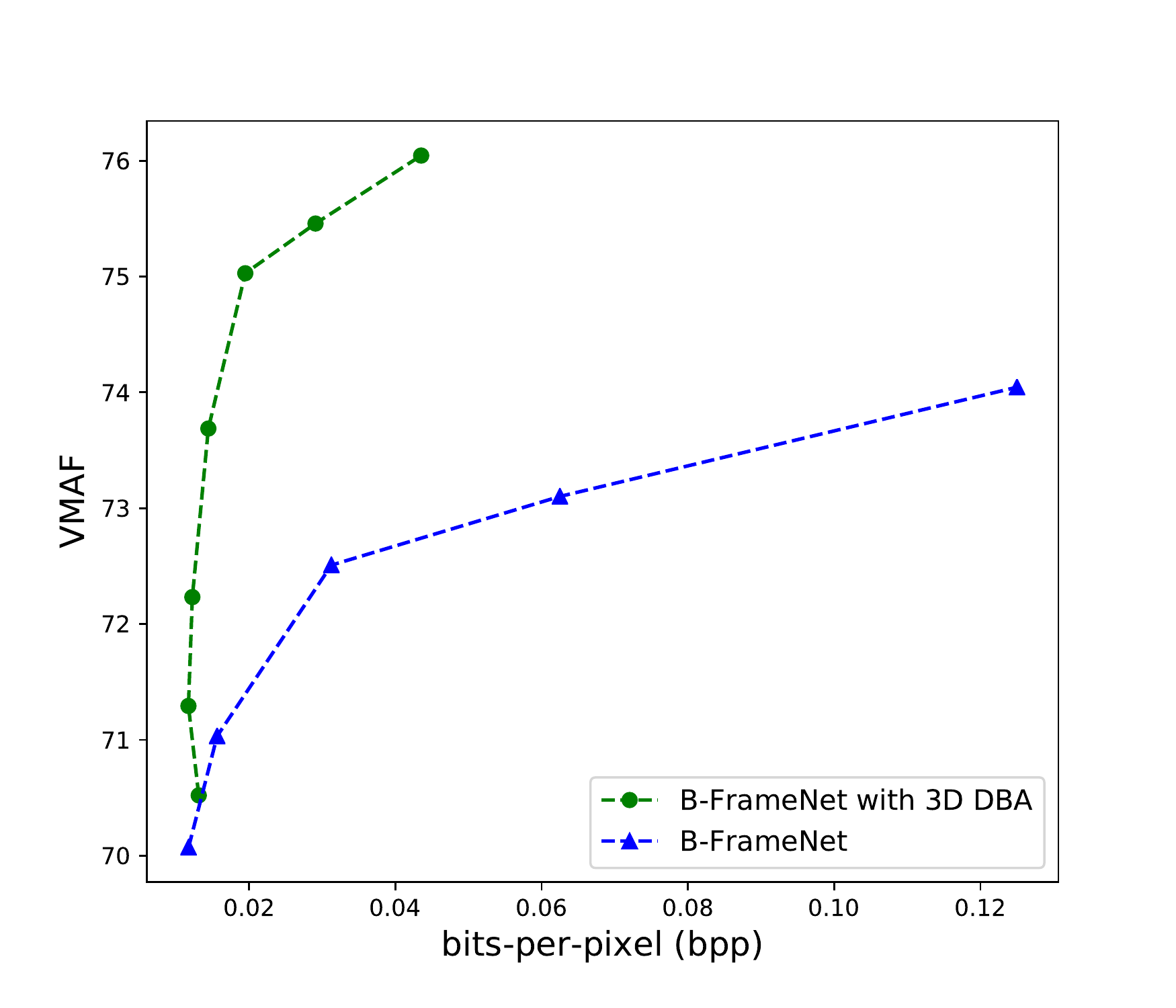}
	}
	\caption{B-FrameNet with and without 3D dynamic bit assignment (3D DBA).}
\label{fig:dba_bf}
\end{figure}

To establish whether 3D dynamic bit assignment aids compression efficiency,
we train B-FrameNet to vary its bitrate across space-time 
(Sections~\ref{sec:dba_arch} and~\ref{sec:dba_train}).
We train several instantiations of B-FrameNet with different bit channel depths, 
$C_{\textrm{bnd}}=\{6, 8, 16, 32, 64\}$, 
to assess the impact of learned dynamic bit assignment at different operating points.
The compression curves in Figure~\ref{fig:dba_bf} show that optimising a model 
for 3D dynamic bit assignment substantially improves 
its PSNR, SSIM and VMAF scores across all bitrates, 
even with the additional overhead of transmitting 
the binarised importance map, $Q_L(B_{\textrm{map}})$.

%
%

%

\subsection{Do Spatial Bit Allocations Change Over Time?}
\begin{figure}[!h]
	\center{
		\includegraphics[width=.73\textwidth]{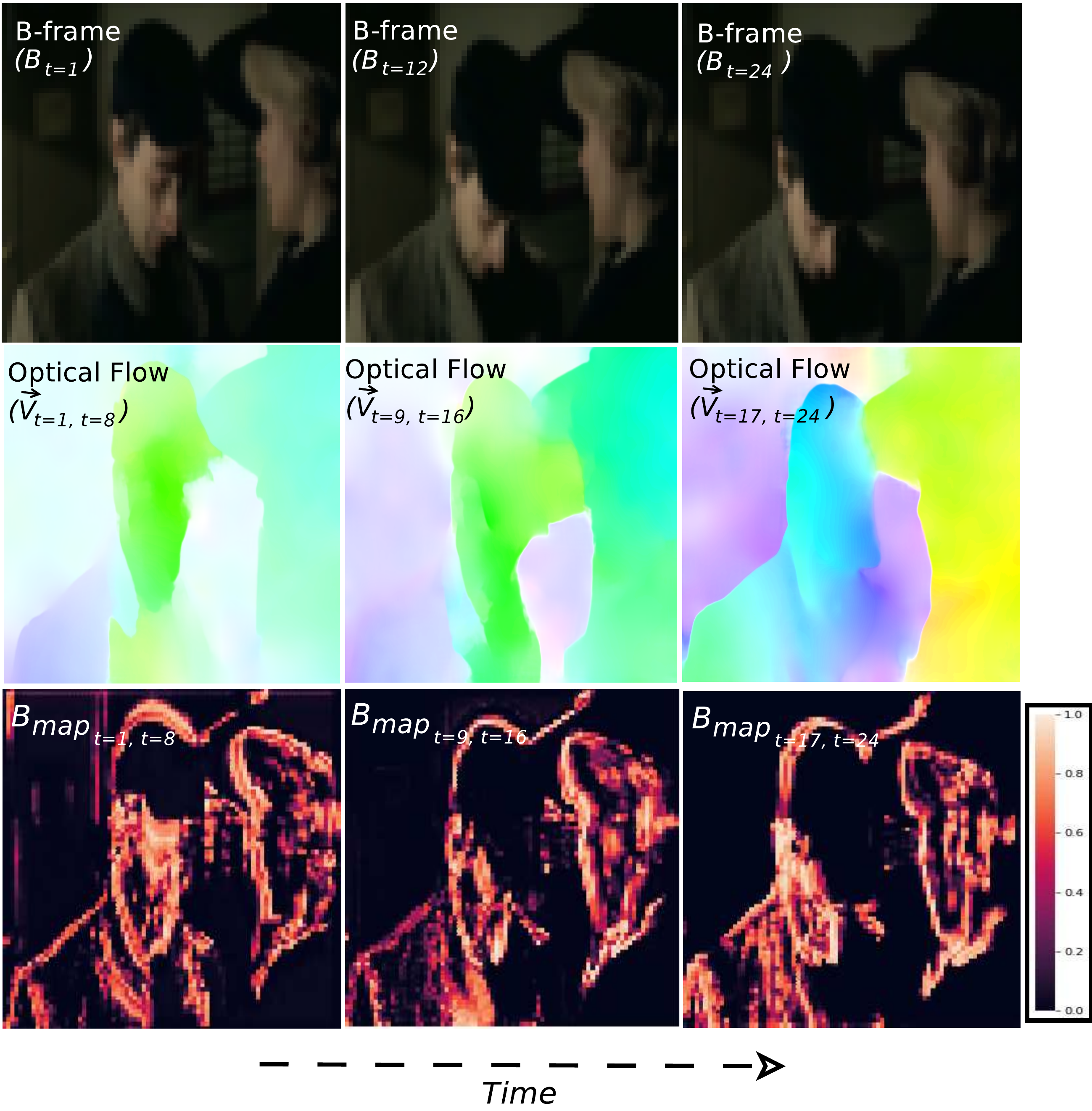}
	}
	\caption{
		B-FrameNet's bit-distribution map, $B_{\textrm{map}}$, 
		compared to optical flow (FlowNet) and input video frames. 
		Brighter regions in $B_{\textrm{map}}$ are allocated higher bitrates 
		and correspond to moving objects.
	}
\label{fig:dba_map}
\end{figure}

Figure~\ref{fig:dba_map} plots B-FrameNet's bit-distribution map 
$B_{\textrm{map}}$ for the given 26-frame input GOP.
Because B-FrameNet compresses both space and time by a factor of around 8,
$B_{\textrm{map}}$ consists of three distinct bit distributions---one per 8 frame interval.
Higher valued regions in $B_{\textrm{map}}$ are brighter 
and correspond to areas encoded with more bits.

The optical flow charts in Figure~\ref{fig:dba_map} are plotted 
in the hue saturation value (HSV) colour space. 
The angular direction of the optical flow vectors is indicated by hue, 
so that vectors pointing in the same direction are coloured the same.
Saturation indicates the magnitude of the vectors,
so vectors with higher magnitudes (moving objects) are less transparent 
and are represented with more intense colours. 
Comparing $B_{\textrm{map}}$ to the optical flow charts in Figure~\ref{fig:dba_map},
we notice qualitatively that more bits are assigned 
to regions containing moving objects 
(brightly coloured regions in the optical flow chart).

$B_{\textrm{map}}$'s spatial bit distribution also changes across time to 
compensate for object displacements caused by motion.
As an aside, for video scenes containing rapid motion or multiple scene changes 
it may help to lessen B-FrameNet's compression factor across time as 
this would yield more frequent updates to $B_{\textrm{map}}$.

%
%


%% file: my_content/eval.tex

\section{Results: Comparing to Conventional Video Compression}
\label{sec:results}

We next compare our learned video compression 
approach to standard codecs.

\subsection{Deep Motion Estimation vs.\ Standard Block Motion Algorithms}
\label{sec:pfbf_vs_bme}
Block-based motion estimation involves finding motion vectors (MVs) that model 
the movement of macroblocks between consecutive video frames.
We compare P-FrameNet and B-FrameNet, $C_{\textrm{bnd}}=8$, 
optimised for 3D dynamic bit assignment to several block-based motion estimation 
algorithms employed by standard video codecs, namely:
%
\begin{itemize}
     \item Exhaustive Search (ES)~\cite{Richards2010}
     \item Three Step Search (TSS)~\cite{Kamble2017}
     \item New Three Step Search (NTSS)~\cite{Li1994}
     \item Simple and Efficient Search (SES)~\cite{Lu1997}
     \item Four Step Search (FSS)~\cite{Po1996}
     \item Diamond Search (DS)~\cite{Zhu2000}
     \item Adaptive Rood Pattern Search (ARPS)~\cite{Nie2002}
\end{itemize}
This evaluation is carried out on videos 
from the VTL dataset~\cite{VTL2000}.
Our models are intended for video prediction only, 
so here we strip down the standard video codecs so 
that only the mechanisms used for inter-frame prediction are compared.
We apply these standard algorithms to $IPPP$ GOP sequences,
so that each macroblock in the currently decoded P-frame is linked to the closest matching 
macroblock region from the preceding frame by way of 
a MV that indicates its relative spatial displacement.

For transmission, 
we binarise each MV's $x$ and $y$ components as well as 
the centre coordinates of the reference macroblock to which it points.
Zero-vectors and overhead bits needed for reshaping are discounted; 
here we only consider bits that effect motion transformation.
Searching all possible pixel locations in the reference frame for 
each predicted macroblock's closest match is computationally expensive, 
especially for high resolution videos. 
Hence, the search area is typically limited to $p=7$ pixels 
around the predicted macroblock's location~\cite{Gall1991}.
We experiment with $\textrm{mb}=8\times8$ and $\textrm{mb}=16\times16$ 
macroblock size parameterisations
of the different algorithms in Tables~\ref{tab:block_motion_eval_16} 
and~\ref{tab:block_motion_eval_8}, respectively.
Smaller macroblocks produce denser MVs resulting in finer motion 
prediction at the cost of a higher bitrate and longer execution time.
In this evaluation all models and algorithms are used to 
predict sixteen $224\times320$ video frames.
Note that for this evaluation we assume uncompressed 
I-frame context is available at the decoder.

Tables~\ref{tab:block_motion_eval_16} and~\ref{tab:block_motion_eval_8} 
show that for fewer bits-per-pixel (bpp), 
P-FrameNet and B-FrameNet's predictions score higher 
in terms of PSNR and VMAF than those produced by the block-matching algorithms.
SSIM scores are comparable, but our models use at least 
23\% and 88\% fewer bpp than the block-matching algorithms in 
Tables~\ref{tab:block_motion_eval_16} and~\ref{tab:block_motion_eval_8}, respectively.
P-FrameNet and B-FrameNet's encoding and decoding time is faster than 
than that of all the block-based motion estimation algorithms.
This speedup stems from their ability to predict frames in parallel 
without the need for a search-step during encoding.

\begin{table}[h]
\caption{Motion compensation scores for 16 frame video prediction $(\textrm{mb}=16\times16, p=7)$.}
	\begin{center}
		\begin{tabularx}{.9\linewidth}{@{}lCCCCCCCC@{}}
			\toprule
				Model & bpp & PSNR & SSIM & VMAF & Time \tiny{(sec)} \\
			\midrule
				ES    & $0.0108$ & $16.35$ & $0.850$ & $44.64$ & $11.35$ \\
				TSS   & $0.0108$ & $16.37$ & $0.851$ & $44.81$ & $1.53$ \\
				NTSS  & $0.0107$ & $16.34$ & $0.851$ & $44.76$ & $1.18$ \\
				SES   & $0.0068$ & $15.80$ & $0.846$ & $40.94$ & $0.96$ \\
				FSS   & $0.0077$ & $16.29$ & $0.852$ & $44.77$ & $1.01$ \\
				DS    & $0.0100$ & $15.70$ & $0.816$ & $37.90$ & $0.77$ \\
				ARPS  & $0.0097$ & $15.66$ & $0.816$ & $37.89$ & $0.63$ \\
				P-FrameNet & $0.0052$ & $28.89$ & $0.829$ & $65.66$ & $0.28$ \\
				B-FrameNet & $\mathbf{0.0038}$ & $\mathbf{30.36}$ & $\mathbf{0.859}$ & $\mathbf{71.19}$ & $\mathbf{0.28}$ \\	
			\bottomrule
		\end{tabularx}
	\end{center}
\label{tab:block_motion_eval_16}
\end{table}

\begin{table}[h]
\caption{Motion compensation scores for 16 frame video prediction $(\textrm{mb}=8\times8, p=7)$.}
	\begin{center}
		\begin{tabularx}{.9\linewidth}{@{}lCCCCCCCC@{}}
			\toprule
				Model & bpp & PSNR & SSIM & VMAF & Time \tiny{(sec)} \\
			\midrule
				ES   & $0.0582$ & $19.45$ & $0.901$ & $62.97$ & $46.32$ \\
				TSS  & $0.0578$ & $19.47$ & $0.901$ & $63.15$ & $5.67$ \\
				NTSS & $0.0568$ & $19.44$ & $\mathbf{0.902}$ & $63.17$ & $4.29$ \\
				SES  & $0.0396$ & $18.80$ & $0.893$ & $57.18$ & $3.71$ \\
				FSS  & $0.0437$ & $19.38$ & $0.901$ & $62.19$ & $3.86$ \\
				DS   & $0.0552$ & $18.58$ & $0.860$ & $53.09$ & $2.82$ \\
				ARPS & $0.0539$ & $18.54$ & $0.861$ & $53.16$ & $2.15$ \\
				P-FrameNet & $0.0052$ & $28.89$ & $0.829$ & $65.66$ & $0.28$ \\
				B-FrameNet & $\mathbf{0.0038}$ & $\mathbf{30.36}$ & $0.859$ & $\mathbf{71.19}$ & $\mathbf{0.28}$ \\	
			\bottomrule
		\end{tabularx}
	\end{center}
	\vspace{-2mm}
\label{tab:block_motion_eval_8}
\end{table}

\subsection{Deep Motion Compression vs.\ Standard Video Codecs}
In Section~\ref{sec:pfbf_vs_bme} we demonstrated that with fewer bits 
our learned binary motion codes are able to express richer motion 
than several block-based motion estimation algorithms.
Standard video codecs improve MV compression via techniques 
not included in the standalone implementation of the algorithms used above.
To reduce bitrate, similar MVs are grouped together and only the Motion Vector Difference (MVD) 
between each vector and a Motion Vector Predictor (MVP) is transmitted~\cite{Yang2010}.
MVD values are normally lower than those of the original MVs, 
especially for a good choice in MVP,\footnote{
	In H.264, a MVP is selected as the mean of a MV group~\cite{ITU-T2003}.
}
and can be represented with fewer bits.
Standard video codecs also actively adapt their block-motion algorithm's 
macroblock-size and search distance to better suit the content of different video regions.

For these reasons, we now compare P-FrameNet and B-FrameNet to 
the standard video codecs H.264~\cite{ITU-T2003} and H.265~\cite{ITU-T2018}.
FFmpeg is used to compress videos with H.264/5.
We varied each codec's constant rate factor (CRF); 
this aims to achieve a constant quality across 
all video frames using as few bits as possible.

Both P-FrameNet and B-FrameNet are intended to provide inter-frame prediction 
of P/B-frames as part of a larger video compression system.
To compress I-frames we adopt H.264's intra-frame codec, 
which is similar to that of H.265~\cite{ITU-T2003,ITU-T2018}.
Any image codec can be used for I-frame compression, 
and more powerful deep image codecs do exist~\cite{Johnston2018,Balle2018,Lee2019},
but to keep this evaluation fair we include P-FrameNet and B-FrameNet 
into an existing video codec in order to investigate the effects 
of including learned motion prediction in isolation.
Since all of the video codecs being evaluated share H.264's I-frame codec,
any compressive gains stem mainly from improved inter-frame coding.
We vary the quality (bit allocation) of the I-frames to gauge 
P-FrameNet and B-FrameNet's performance across different operating points.
For this evaluation, we train $C_{\textrm{bnd}}=8$ versions of P-FrameNet and B-FrameNet 
that have been optimised for 3D dynamic bit assignment.
Bear in mind that unlike H.264 and H.265 our learned binary motion codes 
and overhead bits do not undergo any form of entropy coding.

\subsubsection{P-FrameNet vs.\ Standard Video Codecs}
We plot rate-distortion curves for P-FrameNet, H.264, and H.265, 
based on their respective compression of 17-frame $64\times64$ videos clips 
sampled from the VTL dataset~\cite{VTL2000}.
Each clip consists of a single I-frame followed by sixteen referencing frames.
We allow the standard codecs to decide on their own whether to 
assign referencing frames as P or B (or a mixture of both), 
so that they can achieve their best possible compression.
Quality scores are calculated on and averaged across 
the 17-frame video reconstructions.

\begin{figure}[!h]
\vspace{-7mm}
	\centering
	\subfigure[PSNR]{
		\includegraphics[width=.39\textwidth]{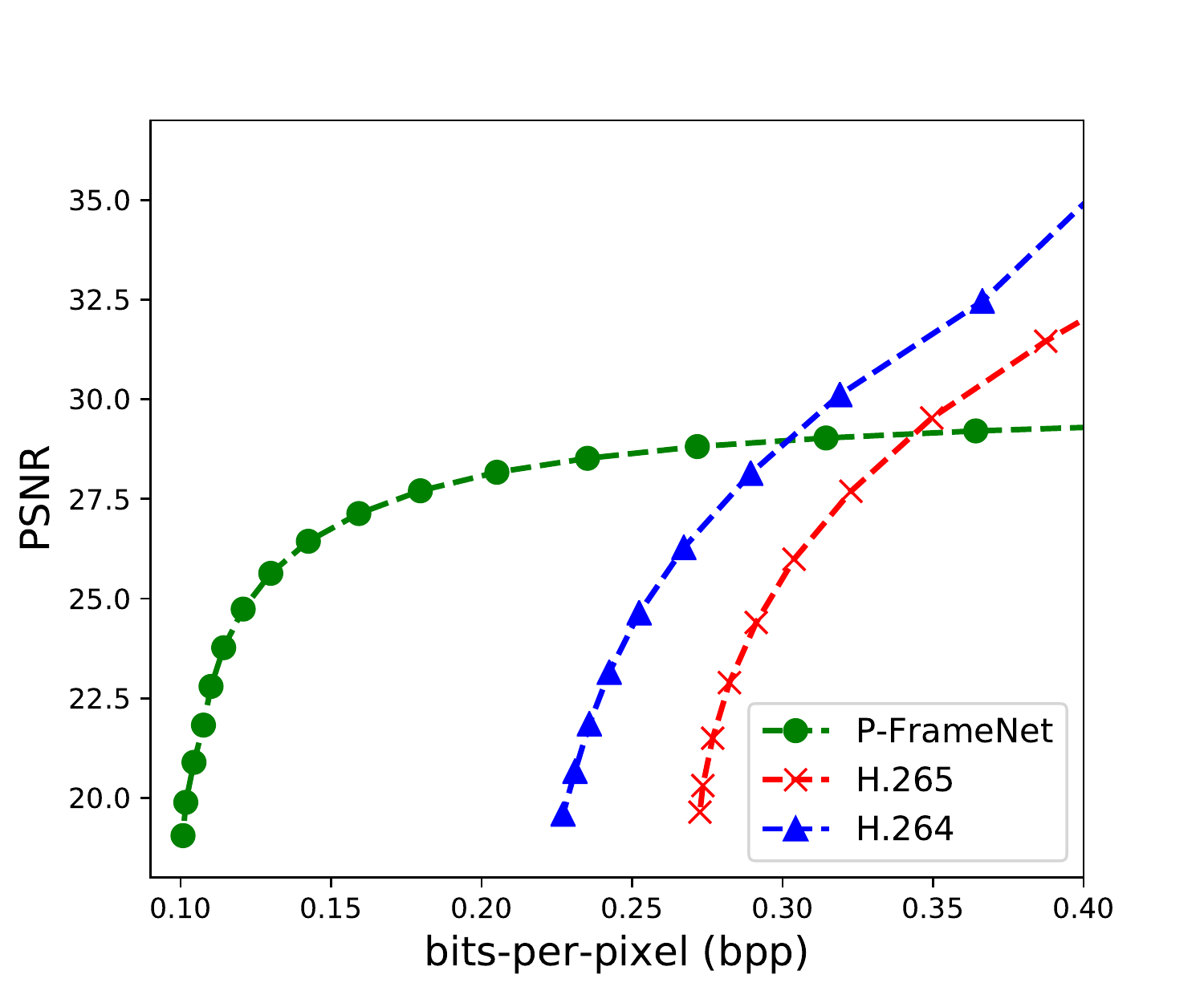}
		\label{fig:pf_vs_std_psnr}
	}
	\subfigure[SSIM]{
		\includegraphics[width=.39\textwidth]{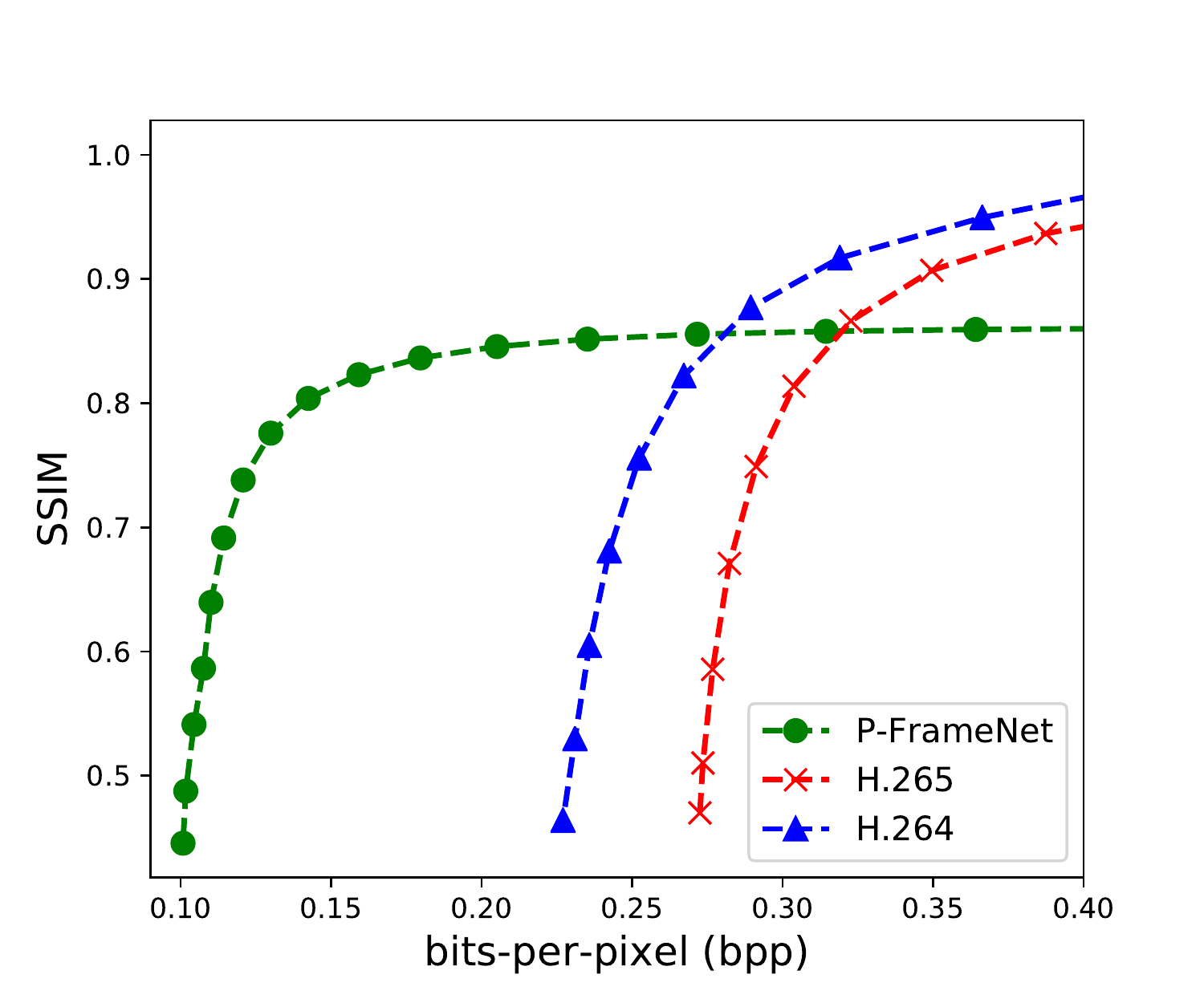}
		\label{fig:pf_vs_std_ssim}
	}
	\subfigure[VMAF]{
		\includegraphics[width=.39\textwidth]{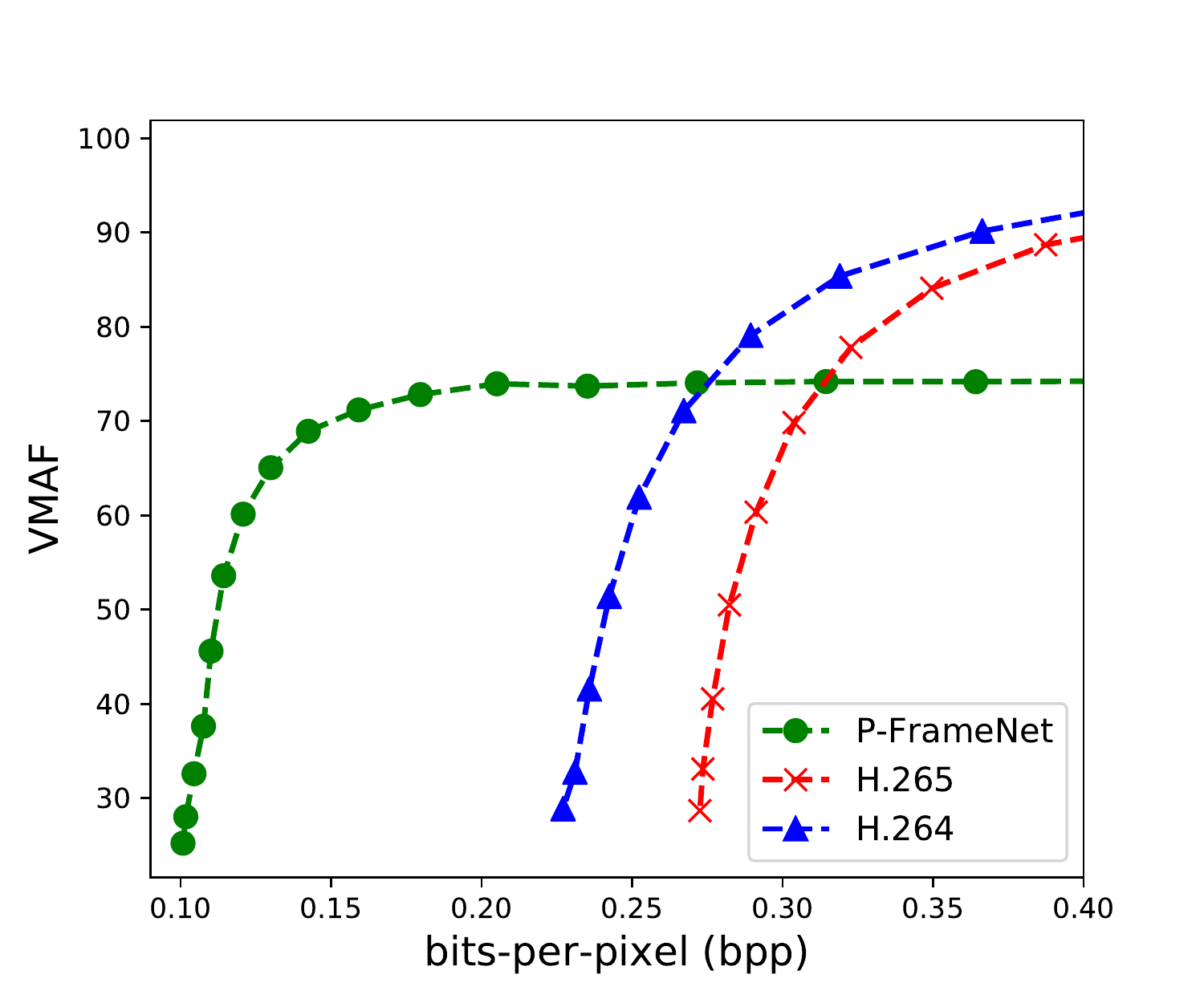}
		\label{fig:pf_vs_std_vmaf}
	}
	\caption{P-FrameNet vs.\ standard video codecs rate-distortion curves.}
	\vspace{-5mm}
\label{fig:pf_vs_std_cc}
\end{figure}

Figure~\ref{fig:pf_vs_std_cc} reveals that P-FrameNet outperforms both H.264 and H.265 
in terms of PSNR, SSIM and VMAF at low-bitrates.
The standard video codecs fair better at higher bit-depths, 
but this is only because P-FrameNet and B-FrameNet do not, as of yet, 
compress residual information to improve the quality of their predictions.
This research concentrated on improving motion estimation and compensation in video compression 
and this is shown by P-FrameNet performance gain at low-bitrates 
where video reconstruction is mostly the result of inter-frame prediction and not residual coding.
 
The link below gives a side-by-side example of P-FrameNet's 
compression compared to that of H.264 and H.265 at a low bitrate. 
The  ground truth video is taken from the wild 
and split into 17-frame GOP groups. 
Each GOP is then encoded and decoded individually. 
The decoded GOPs are then concatenated to re-create the full-length clip.

\begin{center}
    \url{https://youtu.be/LiHLVyIJg_I}
\end{center}
%

%
%

\subsubsection{B-FrameNet vs. Standard Video Codecs}
We compare B-FrameNet's compression efficiency to 
that of the standard H.264 and H.265 video codecs.
Here we compress 18-frame $64\times64$ videos, 
each containing two bounding I-frames and sixteen P/B-frames.

\begin{figure}[!h]
	\centering
	\subfigure[PSNR]{
		\includegraphics[width=.43\textwidth]{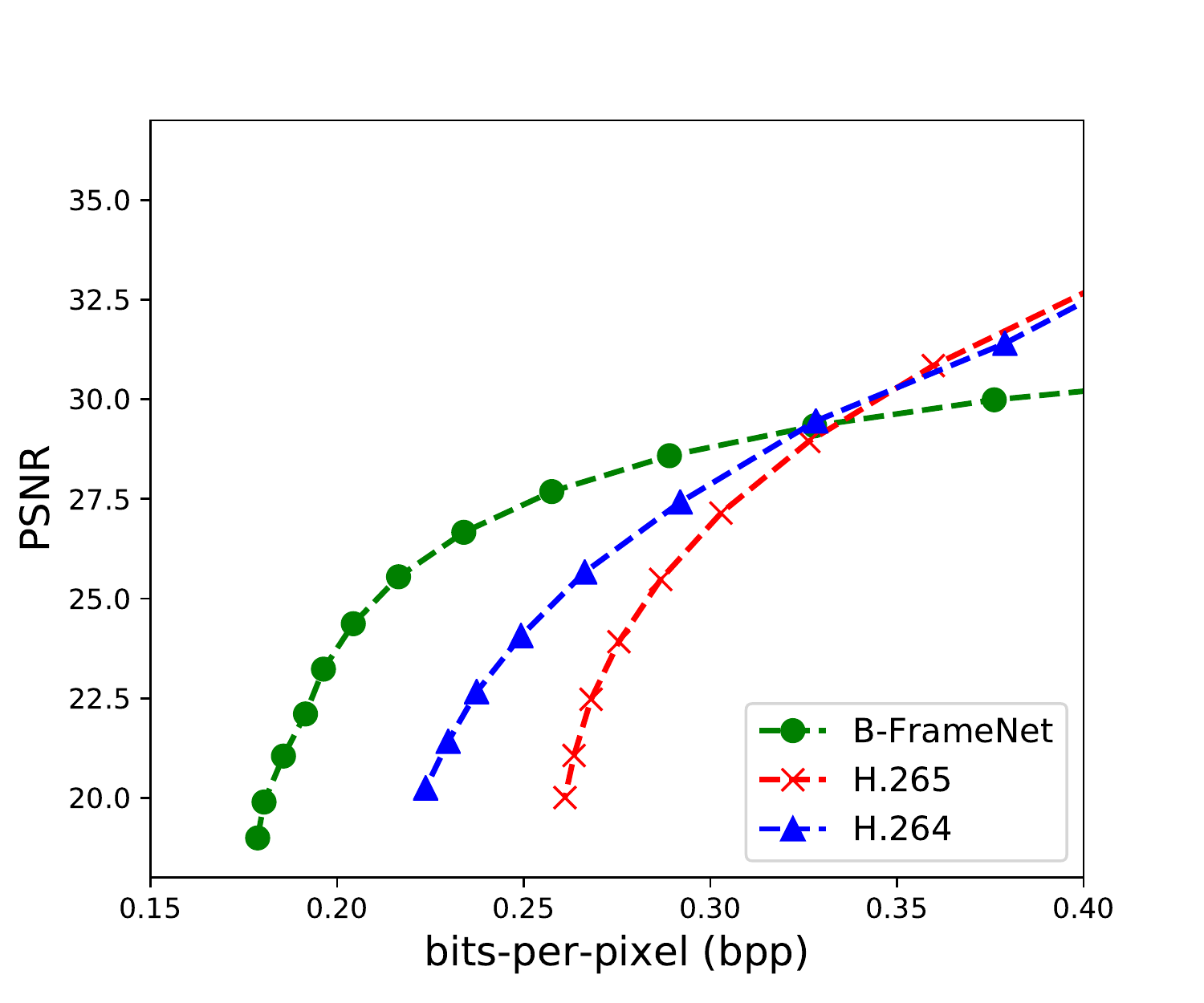}
		\label{fig:bf_vs_std_psnr}
	}
	\subfigure[SSIM]{
		\includegraphics[width=.43\textwidth]{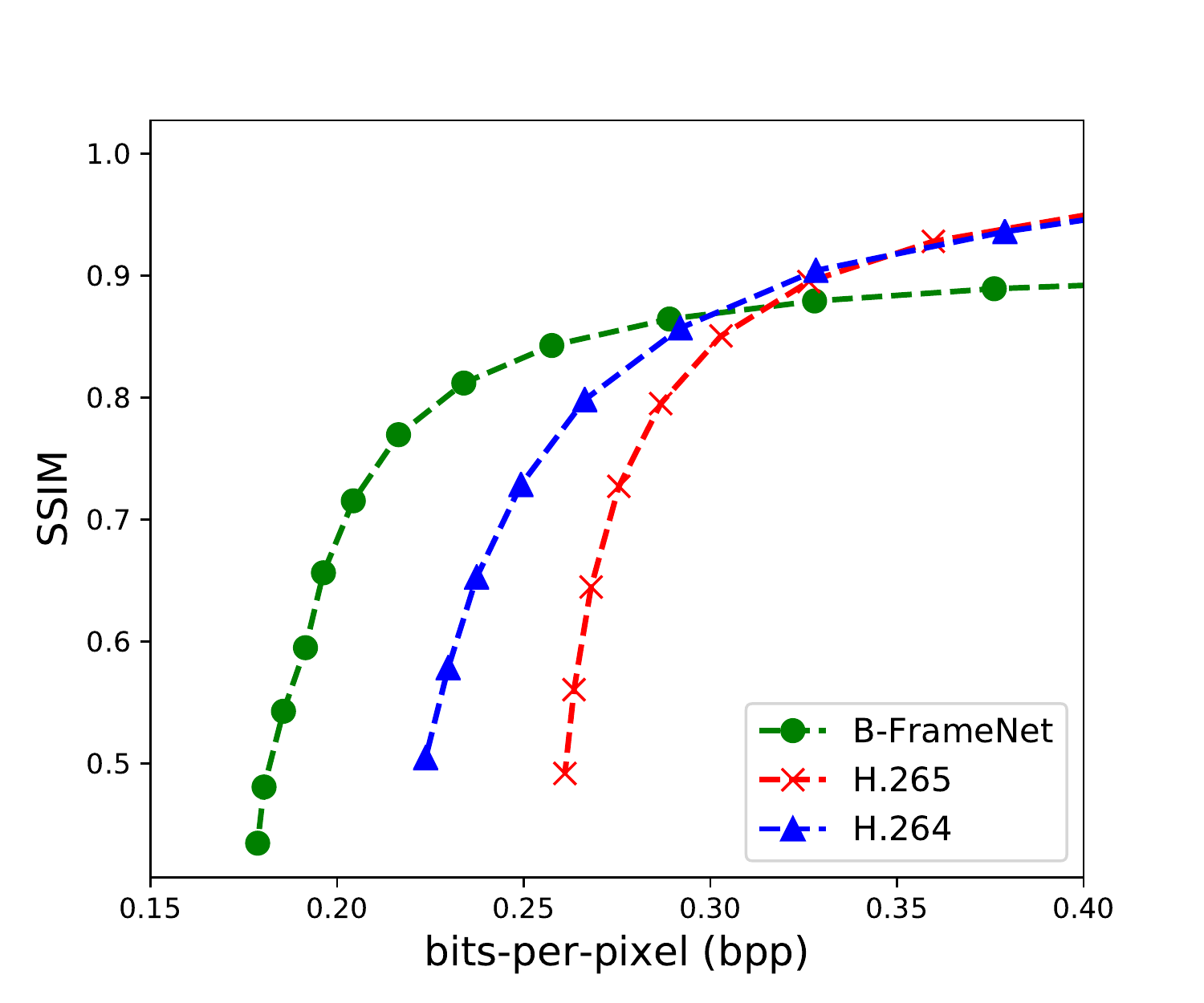}
		\label{fig:bf_vs_std_ssim}
	}
	\subfigure[VMAF]{
		\centering
		\includegraphics[width=.43\textwidth]{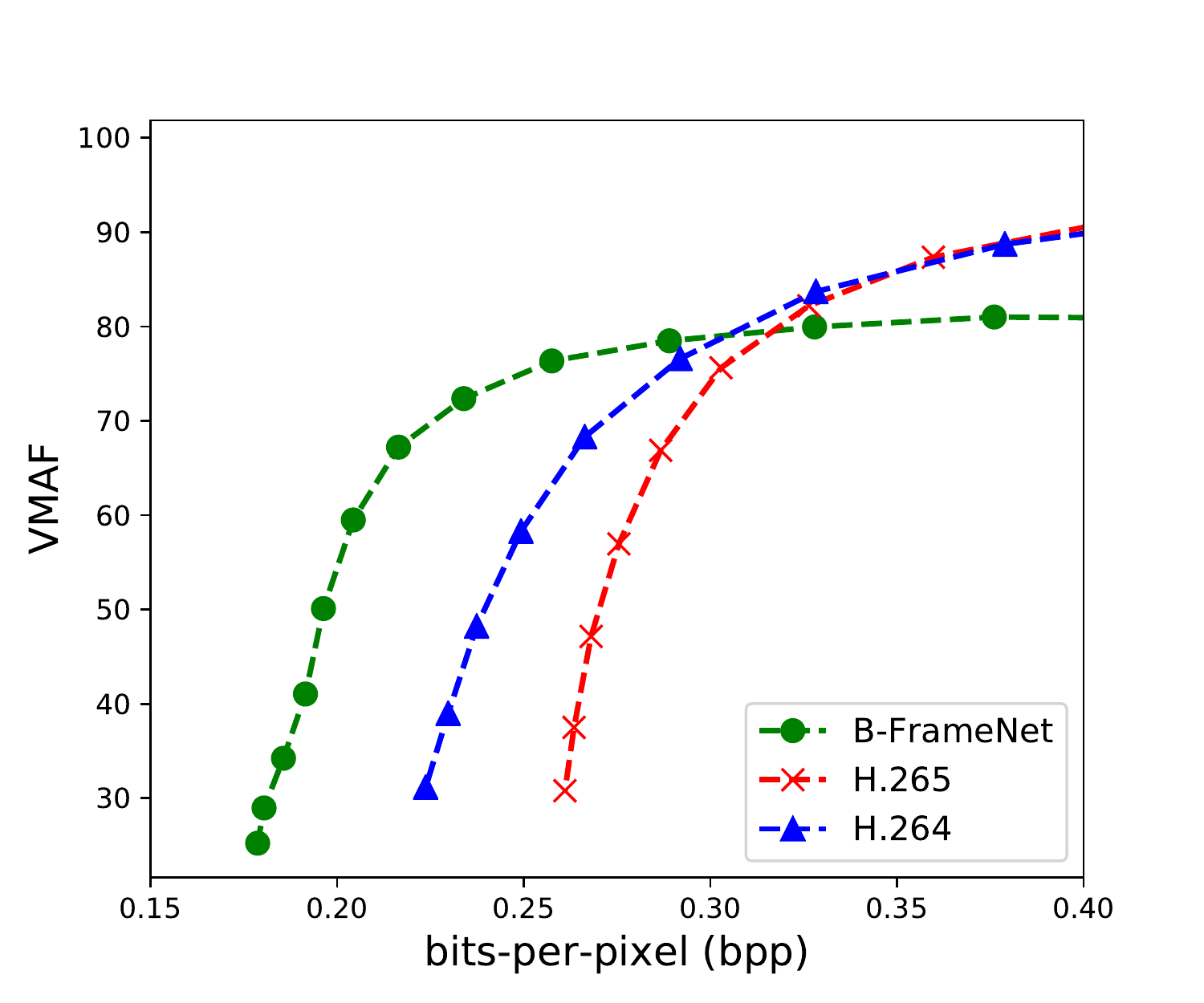}
		\label{fig:bf_vs_std_vmaf}
	}
	\caption{B-FrameNet vs.\ standard video codecs rate-distortion curves.}
\label{fig:bf_vs_std_cc}
\end{figure}

Figure~\ref{fig:bf_vs_std_cc} compares 
the PSNR, SSIM and VMAF rate-distortion curves of the various codecs.
B-FrameNet outperforms all of the standard video codecs at low-bitrates, 
which indicates that it is able to produce higher quality inter-frame predictions.
This claim is further supported by Figure~\ref{fig:bf_vs_std_vid},
which shows the difference in quality between H.264, H.265 
and B-FrameNet's inter-frame predictions.
We only show the middle five predicted video frames for each codec---those 
furthest from the I-frames and most reliant on motion compensation.
It can be seen that B-FrameNet's predictions are qualitatively 
and quantitatively (PSNR, SSIM, VMAF) preferable.

The link below gives a side-by-side example of B-FrameNet's 
compression compared to that of H.264 and H.265 at a low bitrate. 
\begin{center}
    \url{https://youtu.be/nV7mLPwOXTI}
\end{center}
%

%
%

%% file: my_content/conc.tex

%
\begin{figure}[!h]
	\centering
	\subfigure[Ground Truth]{
		\includegraphics[width=.8\textwidth]{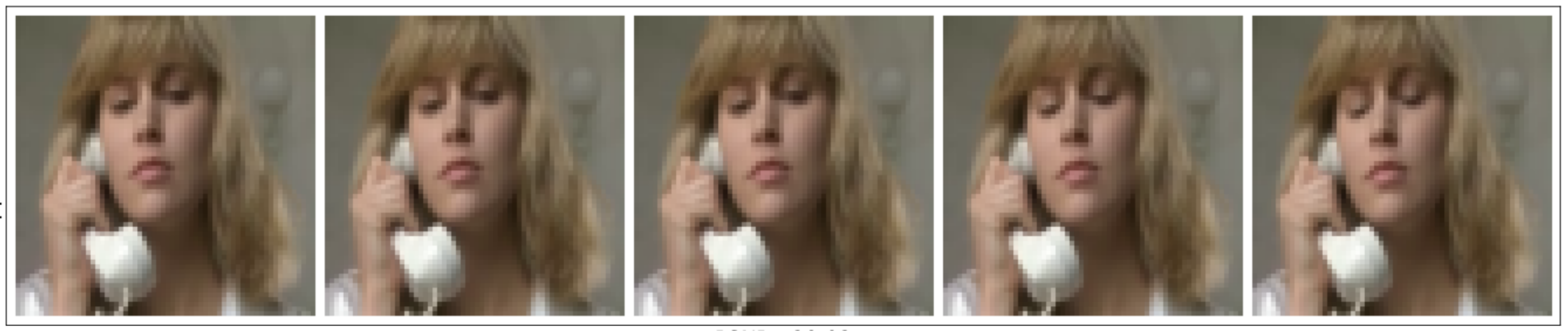}
	}
	\vspace{-2mm}
	\subfigure[B-FrameNet]{
		\includegraphics[width=.8\textwidth]{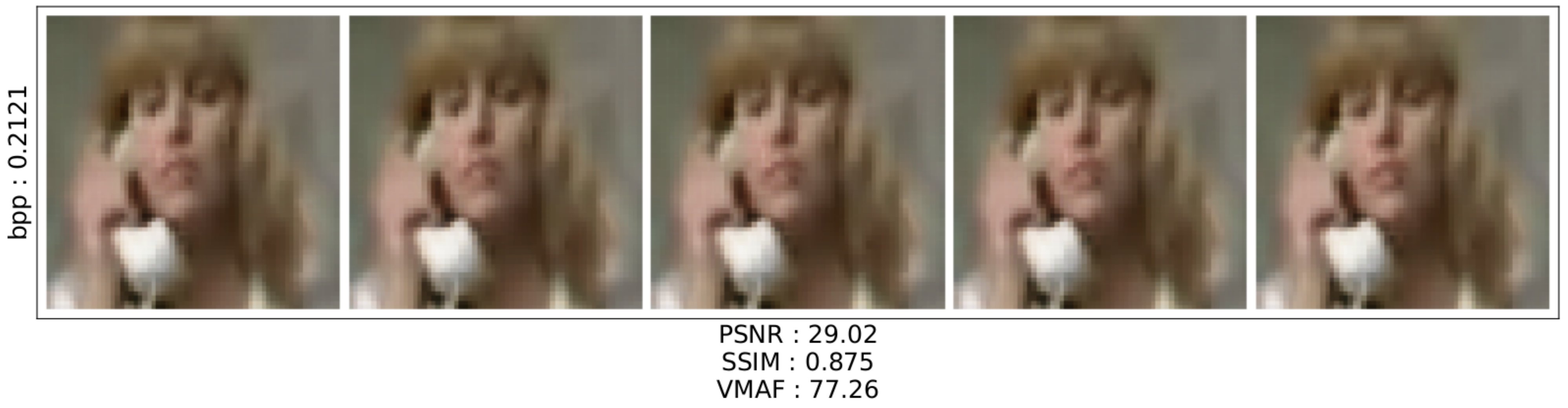}
	}
	\subfigure[H.264]{
		\includegraphics[width=.8\textwidth]{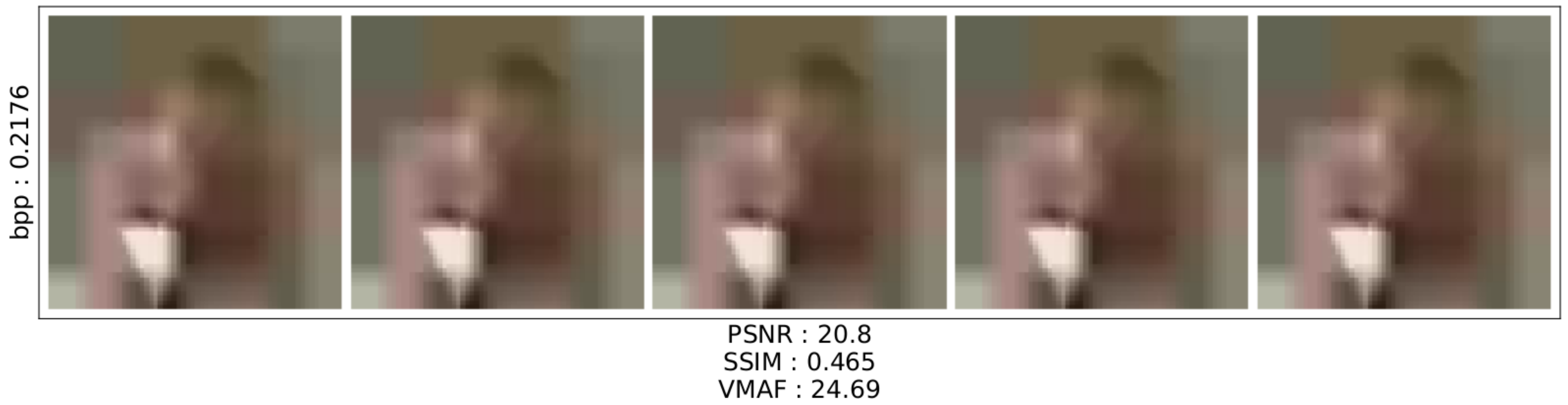}
	}
	\vspace{-1mm}
	\subfigure[H.265]{
		\includegraphics[width=.8\textwidth]{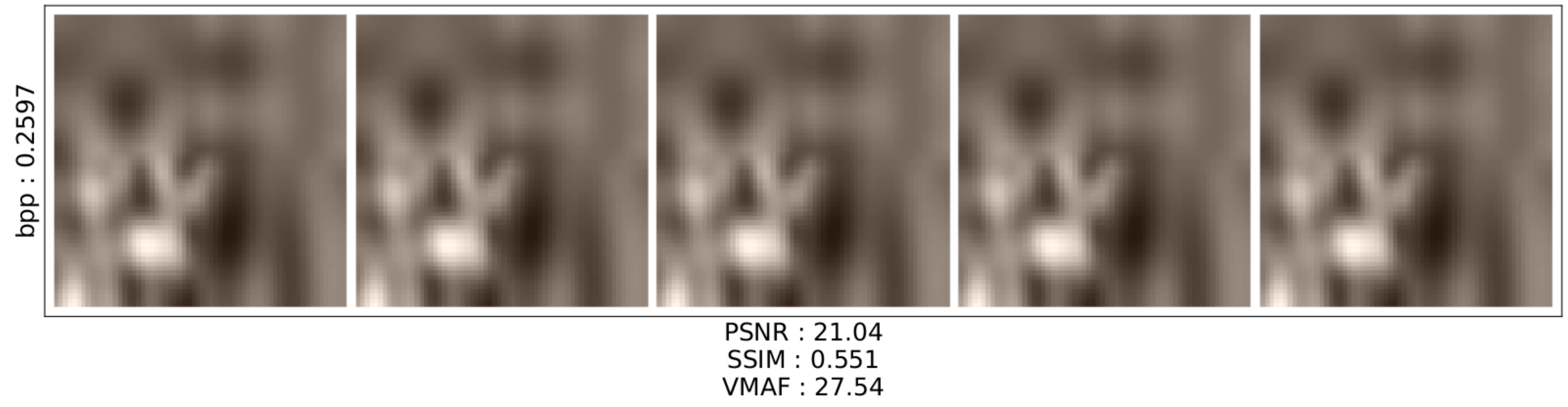}
	}
	\caption{
		B-FrameNet vs.\ standard video codecs inter-frame predictions.
		We only show the five predicted frames midway between the bounding I-frames.  
	}
	\label{fig:bf_vs_std_vid}
\end{figure}

\section{Conclusions} 
We introduced P-FrameNet and B-FrameNet, 
deep motion estimation and compensation networks that can 
replace block-motion algorithms in existing video codecs for improved inter-frame prediction.
In contrast to previously developed video codecs, 
we do not transmit optical flow vectors to guide our video frame predictions. 
Instead, our encoder network learns to identify and compress 
the motion present in a video sequence directly. 
The ensuing binary motion code is used to direct P-FrameNet and B-FrameNet's 
decoder in transforming reference frame content.
This allows for parallel motion compensation that 
predicts more complex motion than flow-based methods.
Leveraging recent work in deep image compression, 
we also train P-FrameNet and B-FrameNet to perform 3D dynamic bit assignment, 
i.e.\ vary their bit allocations through space-time.
We show that this improves compression by focusing bits 
on complicated video regions. 
Experiments show that at a lower bitrate, 
both P-FrameNet and B-FrameNet's inter-frame predictions 
are of a higher quality than those of the standard video codecs, H.264 and H.265.

Apart from porting our inter-frame prediction networks into existing deep video codecs, 
future work will explore replacing flow-based motion estimation in alternative video 
applications (e.g.\ slow-motion) with conditioning on our learned binary motion encodings.


%% file: main.bbl
\begin{thebibliography}{66}
\expandafter\ifx\csname natexlab\endcsname\relax\def\natexlab#1{#1}\fi
\providecommand{\url}[1]{\texttt{#1}}
\providecommand{\href}[2]{#2}
\providecommand{\path}[1]{#1}
\providecommand{\DOIprefix}{doi:}
\providecommand{\ArXivprefix}{arXiv:}
\providecommand{\URLprefix}{URL: }
\providecommand{\Pubmedprefix}{pmid:}
\providecommand{\doi}[1]{\href{http://dx.doi.org/#1}{\path{#1}}}
\providecommand{\Pubmed}[1]{\href{pmid:#1}{\path{#1}}}
\providecommand{\bibinfo}[2]{#2}
\ifx\xfnm\relax \def\xfnm[#1]{\unskip,\space#1}\fi
\bibitem[{Cisco(2017)}]{Cisco2017}
\bibinfo{author}{Cisco}, \bibinfo{title}{{Cisco visual networking index}},
  \bibinfo{howpublished}{available at
  \url{https://www.cisco.com/c/en/us/solutions/collateral/service-provider/visual-networking-index-vni/white-paper-c11-741490.html}},
  \bibinfo{year}{2017}.
\bibitem[{Rippel et~al.(2018)Rippel, Nair, Lew, Branson, Anderson, and
  Bourdev}]{Rippel2018}
\bibinfo{author}{O.~Rippel}, \bibinfo{author}{S.~Nair},
  \bibinfo{author}{C.~Lew}, \bibinfo{author}{S.~Branson},
  \bibinfo{author}{A.~G. Anderson}, \bibinfo{author}{L.~Bourdev},
\newblock \bibinfo{title}{{Learned video compression}},
\newblock \bibinfo{journal}{arXiv preprint arXiv:1811.06981}
  (\bibinfo{year}{2018}).
\bibitem[{Lu et~al.(2019)Lu, Ouyang, Xu, Zhang, Cai, and Gao}]{Lu2019}
\bibinfo{author}{G.~Lu}, \bibinfo{author}{W.~Ouyang}, \bibinfo{author}{D.~Xu},
  \bibinfo{author}{X.~Zhang}, \bibinfo{author}{C.~Cai},
  \bibinfo{author}{Z.~Gao},
\newblock \bibinfo{title}{{DVC: an end-to-end deep video compression
  framework}},
\newblock \bibinfo{journal}{IEEE Computer Society Conference on Computer Vision
  and Pattern Recognition (CVPR)}  (\bibinfo{year}{2019}).
\bibitem[{Nortje et~al.(2019)Nortje, Brink, Engelbrecht, and
  Kamper}]{Nortje2019}
\bibinfo{author}{A.~Nortje}, \bibinfo{author}{W.~Brink}, \bibinfo{author}{H.~A.
  Engelbrecht}, \bibinfo{author}{H.~Kamper},
\newblock \bibinfo{title}{{Improved patch-based image compression with BINet:
  Binary Inpainting Network}},
\newblock \bibinfo{journal}{In Submission}  (\bibinfo{year}{2019}).
\bibitem[{ITU-T H.264(2003)}]{ITU-T2003}
ITU-T H.264, \bibinfo{title}{{H.264 advanced video coding for generic
  audiovisual services}}, \bibinfo{type}{Standard}, International
  Telecommunication Union, \bibinfo{year}{2003}.
\bibitem[{ITU-T H.265(2018)}]{ITU-T2018}
ITU-T H.265, \bibinfo{title}{{H.265 high efficiency video coding}},
  \bibinfo{type}{Standard}, International Telecommunication Union,
  \bibinfo{year}{2018}.
\bibitem[{Le~Gall(1991)}]{Gall1991}
\bibinfo{author}{D.~Le~Gall},
\newblock \bibinfo{title}{{MPEG: a video compression standard for multimedia
  applications}},
\newblock \bibinfo{journal}{Communications of the Association for Computing
  Machinery (ACM)}  (\bibinfo{year}{1991}).
\bibitem[{Richardson(2010)}]{Richards2010}
\bibinfo{author}{I.~E.~G. Richardson}, \bibinfo{title}{{The H.264 advanced
  video compression standard}}, \bibinfo{edition}{2.0} ed.,
  \bibinfo{publisher}{Wiley}, \bibinfo{address}{Chichester, West Sussex},
  \bibinfo{year}{2010}.
\bibitem[{F\"{a}rneback(2003)}]{Farneback2003}
\bibinfo{author}{G.~F\"{a}rneback},
\newblock \bibinfo{title}{{Two-frame motion estimation based on polynomial
  expansion}},
\newblock \bibinfo{journal}{Scandinavian Conference on Image Analysis (SCIA)}
  (\bibinfo{year}{2003}).
\bibitem[{VP9 Version 0.6(2016)}]{Google2016}
VP9 Version 0.6, \bibinfo{title}{{VP9 bitstream and decoding process
  specification}}, \bibinfo{type}{Specification}, Google, \bibinfo{year}{2016}.
\bibitem[{Li et~al.(1994)Li, Zeng, and Lio}]{Li1994}
\bibinfo{author}{R.~Li}, \bibinfo{author}{B.~Zeng}, \bibinfo{author}{M.~L.
  Lio},
\newblock \bibinfo{title}{{A new three-step search algorithm for block motion
  estimation}},
\newblock \bibinfo{journal}{IEEE Transactions on Circuits and Systems for Video
  Technology (TCSVT)}  (\bibinfo{year}{1994}).
\bibitem[{Po and Ma(1996)}]{Po1996}
\bibinfo{author}{L.-M. Po}, \bibinfo{author}{W.-C. Ma},
\newblock \bibinfo{title}{{A novel four-step search algorithm for fast block
  motion estimation}},
\newblock \bibinfo{journal}{IEEE Transactions on Circuits and Systems for Video
  Technology (TCSVT)}  (\bibinfo{year}{1996}).
\bibitem[{Lu and Lio(1997)}]{Lu1997}
\bibinfo{author}{J.~Lu}, \bibinfo{author}{M.~L. Lio},
\newblock \bibinfo{title}{{A simple and efficient search algorithm for
  block-matching motion estimation}},
\newblock \bibinfo{journal}{IEEE Transactions on Circuits and Systems for Video
  Technology (TCSVT)}  (\bibinfo{year}{1997}).
\bibitem[{Zhu and Ma(2000)}]{Zhu2000}
\bibinfo{author}{S.~Zhu}, \bibinfo{author}{K.-K. Ma},
\newblock \bibinfo{title}{{A new diamond search algorithm for fast
  block-matching motion estimation}},
\newblock \bibinfo{journal}{IEEE Transactions on Image Processing (TIP)}
  (\bibinfo{year}{2000}).
\bibitem[{Nie and Ma(2002)}]{Nie2002}
\bibinfo{author}{Y.~Nie}, \bibinfo{author}{K.-K. Ma},
\newblock \bibinfo{title}{{Adaptive rood pattern search for fast block-matching
  motion estimation}},
\newblock \bibinfo{journal}{IEEE Transactions on Image Processing (TIP)}
  (\bibinfo{year}{2002}).
\bibitem[{Haskell et~al.(1997)Haskell, Puri, and Netravali}]{Haskell1997}
\bibinfo{author}{B.~G. Haskell}, \bibinfo{author}{A.~Puri},
  \bibinfo{author}{A.~N. Netravali}, \bibinfo{title}{{Digital video : an
  introduction to MPEG-2}}, \bibinfo{publisher}{Chapman and Hall},
  \bibinfo{year}{1997}.
\bibitem[{De~Brabandere et~al.(2016)De~Brabandere, Jia, Tuytelaars, and
  Van~Gool}]{Jia2016}
\bibinfo{author}{B.~De~Brabandere}, \bibinfo{author}{X.~Jia},
  \bibinfo{author}{T.~Tuytelaars}, \bibinfo{author}{L.~Van~Gool},
\newblock \bibinfo{title}{{Dynamic filter networks}},
\newblock \bibinfo{journal}{Conference on Neural Information Processing Systems
  (NIPS)}  (\bibinfo{year}{2016}).
\bibitem[{Liu et~al.(2017)Liu, Yeh, Tang, Liu, and Agarwala}]{Liu2017}
\bibinfo{author}{Z.~Liu}, \bibinfo{author}{R.~A. Yeh},
  \bibinfo{author}{X.~Tang}, \bibinfo{author}{Y.~Liu},
  \bibinfo{author}{A.~Agarwala},
\newblock \bibinfo{title}{{Video frame synthesis using deep voxel flow}},
\newblock \bibinfo{journal}{International Conference on Computer Vision (ICCV)}
   (\bibinfo{year}{2017}).
\bibitem[{Niklaus et~al.(2017)Niklaus, Mai, and Liu}]{Niklaus2017}
\bibinfo{author}{S.~Niklaus}, \bibinfo{author}{L.~Mai},
  \bibinfo{author}{F.~Liu},
\newblock \bibinfo{title}{{Video frame interpolation via adaptive
  convolution}},
\newblock \bibinfo{journal}{IEEE Computer Society Conference on Computer Vision
  and Pattern Recognition (CVPR)}  (\bibinfo{year}{2017}).
\bibitem[{Niklaus and Liu(2018)}]{Niklaus2018}
\bibinfo{author}{S.~Niklaus}, \bibinfo{author}{F.~Liu},
\newblock \bibinfo{title}{{Context-aware synthesis for video frame
  interpolation}},
\newblock \bibinfo{journal}{IEEE Computer Society Conference on Computer Vision
  and Pattern Recognition (CVPR)}  (\bibinfo{year}{2018}).
\bibitem[{Jiang et~al.(2018)Jiang, Sun, Jampani, Yang, Learned-Miller, and
  Kautz}]{Jiang2018}
\bibinfo{author}{H.~Jiang}, \bibinfo{author}{D.~Sun},
  \bibinfo{author}{V.~Jampani}, \bibinfo{author}{M.-H. Yang},
  \bibinfo{author}{E.~Learned-Miller}, \bibinfo{author}{J.~Kautz},
\newblock \bibinfo{title}{{Super SloMo: high quality estimation of multiple
  intermediate frames for video interpolation}},
\newblock \bibinfo{journal}{IEEE Computer Society Conference on Computer Vision
  and Pattern Recognition (CVPR)}  (\bibinfo{year}{2018}).
\bibitem[{Meyer et~al.(2018)Meyer, Djelouah, McWilliams, Sorkine-Hornung,
  Gross, and Schroers}]{Meyer2018}
\bibinfo{author}{S.~Meyer}, \bibinfo{author}{A.~Djelouah},
  \bibinfo{author}{B.~McWilliams}, \bibinfo{author}{A.~Sorkine-Hornung},
  \bibinfo{author}{M.~Gross}, \bibinfo{author}{C.~Schroers},
\newblock \bibinfo{title}{{PhaseNet for video frame interpolation}},
\newblock \bibinfo{journal}{IEEE Computer Society Conference on Computer Vision
  and Pattern Recognition (CVPR)}  (\bibinfo{year}{2018}).
\bibitem[{Bao et~al.(2019)Bao, Lai, Ma, Zhang, Gao, and Yang}]{Bao2019}
\bibinfo{author}{W.~Bao}, \bibinfo{author}{W.-S. Lai}, \bibinfo{author}{C.~Ma},
  \bibinfo{author}{X.~Zhang}, \bibinfo{author}{Z.~Gao}, \bibinfo{author}{M.-H.
  Yang},
\newblock \bibinfo{title}{{Depth-aware video frame interpolation}},
\newblock \bibinfo{journal}{IEEE Computer Society Conference on Computer Vision
  and Pattern Recognition (CVPR)}  (\bibinfo{year}{2019}).
\bibitem[{Liu et~al.(2019)Liu, Liao, Lin, and Chuang}]{Liu2019}
\bibinfo{author}{Y.-L. Liu}, \bibinfo{author}{Y.-T. Liao},
  \bibinfo{author}{Y.-Y.~L. Lin}, \bibinfo{author}{Y.-Y. Chuang},
\newblock \bibinfo{title}{{Deep video frame interpolation using cyclic frame
  generation}},
\newblock \bibinfo{journal}{Association for the Advancement of Artificial
  Intelligence (AAAI)}  (\bibinfo{year}{2019}).
\bibitem[{Mathieu et~al.(2016)Mathieu, Couprie, and LeCun}]{Mathieu2016}
\bibinfo{author}{M.~Mathieu}, \bibinfo{author}{C.~Couprie},
  \bibinfo{author}{Y.~LeCun},
\newblock \bibinfo{title}{{Deep multi-scale video prediction beyond mean square
  error}},
\newblock \bibinfo{journal}{International Conference on Learning
  Representations (ICLR)}  (\bibinfo{year}{2016}).
\bibitem[{Vondrick et~al.(2016)Vondrick, Pirsiavash, and
  Torralba}]{Vondrick2016}
\bibinfo{author}{C.~Vondrick}, \bibinfo{author}{H.~Pirsiavash},
  \bibinfo{author}{A.~Torralba},
\newblock \bibinfo{title}{{Generating videos with scene dynamics}},
\newblock \bibinfo{journal}{Conference on Neural Information Processing Systems
  (NIPS)}  (\bibinfo{year}{2016}).
\bibitem[{Xue et~al.(2016)Xue, Wu, Bouman, and Freeman}]{Xue2016}
\bibinfo{author}{T.~Xue}, \bibinfo{author}{J.~Wu}, \bibinfo{author}{K.~L.
  Bouman}, \bibinfo{author}{W.~T. Freeman},
\newblock \bibinfo{title}{{Visual dynamics: probabilistic future frame
  synthesis via cross convolutional networks}},
\newblock \bibinfo{journal}{Conference on Neural Information Processing Systems
  (NIPS)}  (\bibinfo{year}{2016}).
\bibitem[{Finn et~al.(2016)Finn, Goodfellow, and Levine}]{Finn2016}
\bibinfo{author}{C.~Finn}, \bibinfo{author}{I.~Goodfellow},
  \bibinfo{author}{S.~Levine},
\newblock \bibinfo{title}{{Unsupervised learning for physical interaction
  through video prediction}},
\newblock \bibinfo{journal}{Conference on Neural Information Processing Systems
  (NIPS)}  (\bibinfo{year}{2016}).
\bibitem[{Wu et~al.(2018)Wu, Singhal, and Krahenbuhl}]{Wu2018}
\bibinfo{author}{C.-Y. Wu}, \bibinfo{author}{N.~Singhal},
  \bibinfo{author}{P.~Krahenbuhl},
\newblock \bibinfo{title}{{Video compression through image interpolation}},
\newblock \bibinfo{journal}{European Conference on Computer Vision (ECCV)}
  (\bibinfo{year}{2018}).
\bibitem[{Chen et~al.(2018)Chen, He, Jin, and Wu}]{Chen2018}
\bibinfo{author}{Z.~Chen}, \bibinfo{author}{T.~He}, \bibinfo{author}{X.~Jin},
  \bibinfo{author}{F.~Wu},
\newblock \bibinfo{title}{{Learning for video compression}},
\newblock \bibinfo{journal}{IEEE Transactions on Circuits and Systems for Video
  Technology (TCSVT)}  (\bibinfo{year}{2018}).
\bibitem[{Habibian et~al.(2019)Habibian, van Rozendaal, Tomczak, and
  Cohen}]{Habibian2019}
\bibinfo{author}{A.~Habibian}, \bibinfo{author}{T.~van Rozendaal},
  \bibinfo{author}{J.~M. Tomczak}, \bibinfo{author}{T.~S. Cohen},
\newblock \bibinfo{title}{{Video compression with rate-distortion
  autoencoders}},
\newblock \bibinfo{journal}{International Conference on Computer Vision (ICCV)}
   (\bibinfo{year}{2019}).
\bibitem[{Cheng et~al.(2019)Cheng, Sun, Takeuchi, and Katto}]{Cheng2019}
\bibinfo{author}{Z.~Cheng}, \bibinfo{author}{H.~Sun},
  \bibinfo{author}{M.~Takeuchi}, \bibinfo{author}{J.~Katto},
\newblock \bibinfo{title}{{Learning image and video compression through
  spatial-temporal energy compaction}},
\newblock \bibinfo{journal}{IEEE Computer Society Conference on Computer Vision
  and Pattern Recognition (CVPR)}  (\bibinfo{year}{2019}).
\bibitem[{Horn and Schunck(1981)}]{Horn1981}
\bibinfo{author}{B.~K. Horn}, \bibinfo{author}{B.~G. Schunck},
\newblock \bibinfo{title}{{Determining optical flow}},
\newblock \bibinfo{journal}{Artificial Intelligence (AI)}
  (\bibinfo{year}{1981}).
\bibitem[{Ho et~al.(2019)Ho, Cho, Peng, and Jin}]{Ho2019}
\bibinfo{author}{Y.-H. Ho}, \bibinfo{author}{C.-Y. Cho}, \bibinfo{author}{W.-H.
  Peng}, \bibinfo{author}{G.-L. Jin},
\newblock \bibinfo{title}{{SME-Net: sparse motion estimation for parametric
  video prediction through reinforcement learning}},
\newblock \bibinfo{journal}{International Conference on Computer Vision (ICCV)}
   (\bibinfo{year}{2019}).
\bibitem[{Johnston et~al.(2018)Johnston, Vincent, Minnen, Covell, Singh,
  Chinen, {Jin Hwang}, Shor, and Toderici}]{Johnston2018}
\bibinfo{author}{N.~Johnston}, \bibinfo{author}{D.~Vincent},
  \bibinfo{author}{D.~Minnen}, \bibinfo{author}{M.~Covell},
  \bibinfo{author}{S.~Singh}, \bibinfo{author}{T.~Chinen},
  \bibinfo{author}{S.~{Jin Hwang}}, \bibinfo{author}{J.~Shor},
  \bibinfo{author}{G.~Toderici},
\newblock \bibinfo{title}{{Improved lossy image compression with priming and
  spatially adaptive bit rates for recurrent networks}},
\newblock \bibinfo{journal}{IEEE Computer Society Conference on Computer Vision
  and Pattern Recognition (CVPR)}  (\bibinfo{year}{2018}).
\bibitem[{Minnen et~al.(2018)Minnen, Ball\'{e}, and Toderici}]{Minnen2018}
\bibinfo{author}{D.~Minnen}, \bibinfo{author}{J.~Ball\'{e}},
  \bibinfo{author}{G.~Toderici},
\newblock \bibinfo{title}{{Joint autoregressive and hierarchical priors for
  learned image compression}},
\newblock \bibinfo{journal}{Conference on Neural Information Processing Systems
  (NIPS)}  (\bibinfo{year}{2018}).
\bibitem[{Li et~al.(2018)Li, Zuo, Gu, Zhao, and Zhang}]{Li2018}
\bibinfo{author}{M.~Li}, \bibinfo{author}{W.~Zuo}, \bibinfo{author}{S.~Gu},
  \bibinfo{author}{D.~Zhao}, \bibinfo{author}{D.~Zhang},
\newblock \bibinfo{title}{{Learning convolutional networks for content-weighted
  image compression}},
\newblock \bibinfo{journal}{IEEE Computer Society Conference on Computer Vision
  and Pattern Recognition (CVPR)}  (\bibinfo{year}{2018}).
\bibitem[{Ball{\'{e}} et~al.(2018)Ball{\'{e}}, Minnen, Singh, Hwang, and
  Johnston}]{Balle2018}
\bibinfo{author}{J.~Ball{\'{e}}}, \bibinfo{author}{D.~Minnen},
  \bibinfo{author}{S.~Singh}, \bibinfo{author}{S.~J. Hwang},
  \bibinfo{author}{N.~Johnston},
\newblock \bibinfo{title}{{Variational image compression with a scale
  hyperprior}},
\newblock \bibinfo{journal}{arXiv preprint arXiv:1802.01436}
  (\bibinfo{year}{2018}).
\bibitem[{Jooyoung~Lee and Beack(2019)}]{Lee2019}
\bibinfo{author}{S.~C. Jooyoung~Lee}, \bibinfo{author}{S.-K. Beack},
\newblock \bibinfo{title}{{Context-adaptive entropy model for end-to-end
  optimized image compression}},
\newblock \bibinfo{journal}{International Conference on Learning
  Representations (ICLR)}  (\bibinfo{year}{2019}).
\bibitem[{Han et~al.(2018)Han, Lombardo, Schroers, and Mandt}]{Han2018}
\bibinfo{author}{J.~Han}, \bibinfo{author}{S.~Lombardo},
  \bibinfo{author}{C.~Schroers}, \bibinfo{author}{S.~Mandt},
\newblock \bibinfo{title}{{Deep probabilistic video compression}},
\newblock \bibinfo{journal}{arXiv preprint arXiv:1810.02845}
  (\bibinfo{year}{2018}).
\bibitem[{Toderici et~al.(2015)Toderici, O'Malley, Hwang, Vincent, Minnen,
  Baluja, Covell, and Sukthankar}]{Toderici2015}
\bibinfo{author}{G.~Toderici}, \bibinfo{author}{S.~M. O'Malley},
  \bibinfo{author}{S.~J. Hwang}, \bibinfo{author}{D.~Vincent},
  \bibinfo{author}{D.~Minnen}, \bibinfo{author}{S.~Baluja},
  \bibinfo{author}{M.~Covell}, \bibinfo{author}{R.~Sukthankar},
\newblock \bibinfo{title}{{Variable rate image compression with recurrent
  neural networks}},
\newblock \bibinfo{journal}{International Conference on Learning
  Representations (ICLR)}  (\bibinfo{year}{2015}).
\bibitem[{Raiko et~al.(2014)Raiko, Berglund, Alain, and Dinh}]{Raiko2014}
\bibinfo{author}{T.~Raiko}, \bibinfo{author}{M.~Berglund},
  \bibinfo{author}{G.~Alain}, \bibinfo{author}{L.~Dinh},
\newblock \bibinfo{title}{{Techniques for learning binary stochastic
  feedforward neural networks}},
\newblock \bibinfo{journal}{arXiv preprint arXiv:1406.2989}
  (\bibinfo{year}{2014}).
\bibitem[{Yu and Koltun(2016)}]{Yu2016}
\bibinfo{author}{F.~Yu}, \bibinfo{author}{V.~Koltun},
\newblock \bibinfo{title}{{Multi-scale context aggregation by dilated
  convolutions}},
\newblock \bibinfo{journal}{International Conference on Learning
  Representations (ICLR)}  (\bibinfo{year}{2016}).
\bibitem[{Baig et~al.(2017)Baig, Koltun, and Torresani}]{Baig2017}
\bibinfo{author}{M.~H. Baig}, \bibinfo{author}{V.~Koltun},
  \bibinfo{author}{L.~Torresani},
\newblock \bibinfo{title}{{Learning to inpaint for image compression}},
\newblock \bibinfo{journal}{Conference on Neural Information Processing Systems
  (NIPS)}  (\bibinfo{year}{2017}).
\bibitem[{Rippel and Bourdev(2017)}]{Rippel2017}
\bibinfo{author}{O.~Rippel}, \bibinfo{author}{L.~Bourdev},
\newblock \bibinfo{title}{{Real-time adaptive image compression}},
\newblock \bibinfo{journal}{arXiv preprint arXiv:1705.05823}
  (\bibinfo{year}{2017}).
\bibitem[{Lowe(2004)}]{Lowe2004}
\bibinfo{author}{D.~G. Lowe},
\newblock \bibinfo{title}{{Distinctive image features from scale-invariant
  keypoints}},
\newblock \bibinfo{journal}{International Journal of Computer Vision (IJCV)}
  (\bibinfo{year}{2004}).
\bibitem[{Ronneberger et~al.(2015)Ronneberger, Fischer, and
  Brox}]{Ronneberger2015}
\bibinfo{author}{O.~Ronneberger}, \bibinfo{author}{P.~Fischer},
  \bibinfo{author}{T.~Brox},
\newblock \bibinfo{title}{{U-Net: convolutional networks for biomedical image
  segmentation}},
\newblock \bibinfo{journal}{International Conference on Medical Image Computing
  and Computer-Assisted Intervention (MICCAI)}  (\bibinfo{year}{2015}).
\bibitem[{Shi et~al.(2016)Shi, Caballero, Huszár, Totz, Aitken, Bishop,
  Rueckert, and Wang}]{Shi2016}
\bibinfo{author}{W.~Shi}, \bibinfo{author}{J.~Caballero},
  \bibinfo{author}{F.~Huszár}, \bibinfo{author}{J.~Totz},
  \bibinfo{author}{A.~P. Aitken}, \bibinfo{author}{R.~Bishop},
  \bibinfo{author}{D.~Rueckert}, \bibinfo{author}{Z.~Wang},
\newblock \bibinfo{title}{{Real-time single image and video super-resolution
  using an efficient sub-pixel convolutional neural network}},
\newblock \bibinfo{journal}{IEEE Computer Society Conference on Computer Vision
  and Pattern Recognition (CVPR)}  (\bibinfo{year}{2016}).
\bibitem[{Lathi and Ding(2018)}]{Lathi2018}
\bibinfo{author}{B.~P. Lathi}, \bibinfo{author}{Z.~Ding},
  \bibinfo{title}{{Modern digital and analog communication systems}},
  \bibinfo{edition}{4.0} ed., \bibinfo{publisher}{Oxford University Press},
  \bibinfo{year}{2018}.
\bibitem[{Gautama and Van~Hulle(2002)}]{Gautama2002}
\bibinfo{author}{T.~Gautama}, \bibinfo{author}{M.~M. Van~Hulle},
\newblock \bibinfo{title}{{A phase-based approach to the estimation of the
  optical flow field using spatial filtering}},
\newblock \bibinfo{journal}{IEEE Transactions on Neural Networks (TNN)}
  (\bibinfo{year}{2002}).
\bibitem[{Fischer et~al.(2015)Fischer, Dosovitskiy, Ilg, Häusser, Hazırbaş,
  Golkov, van~der Smagt, Cremers, and Brox}]{Fischer2015}
\bibinfo{author}{P.~Fischer}, \bibinfo{author}{A.~Dosovitskiy},
  \bibinfo{author}{E.~Ilg}, \bibinfo{author}{P.~Häusser},
  \bibinfo{author}{C.~Hazırbaş}, \bibinfo{author}{V.~Golkov},
  \bibinfo{author}{P.~van~der Smagt}, \bibinfo{author}{D.~Cremers},
  \bibinfo{author}{T.~Brox},
\newblock \bibinfo{title}{{FlowNet: learning optical flow with convolutional
  networks}},
\newblock \bibinfo{journal}{International Conference on Computer Vision (ICCV)}
   (\bibinfo{year}{2015}).
\bibitem[{Ilg et~al.(2016)Ilg, Mayer, Saikia, Keuper, Dosovitskiy, and
  Brox}]{Ilg2016}
\bibinfo{author}{E.~Ilg}, \bibinfo{author}{N.~Mayer},
  \bibinfo{author}{T.~Saikia}, \bibinfo{author}{M.~Keuper},
  \bibinfo{author}{A.~Dosovitskiy}, \bibinfo{author}{T.~Brox},
\newblock \bibinfo{title}{{FlowNet 2.0: evolution of optical flow estimation
  with deep networks}},
\newblock \bibinfo{journal}{IEEE Computer Society Conference on Computer Vision
  and Pattern Recognition (CVPR)}  (\bibinfo{year}{2016}).
\bibitem[{Zweig and Wolf(2016)}]{Zweig2016}
\bibinfo{author}{S.~Zweig}, \bibinfo{author}{L.~Wolf},
\newblock \bibinfo{title}{{InterpoNet, A brain inspired neural network for
  optical flow dense interpolation}},
\newblock \bibinfo{journal}{IEEE Computer Society Conference on Computer Vision
  and Pattern Recognition (CVPR)}  (\bibinfo{year}{2016}).
\bibitem[{Ranjan and Black(2017)}]{Ranjan2017}
\bibinfo{author}{A.~Ranjan}, \bibinfo{author}{M.~J. Black},
\newblock \bibinfo{title}{{Optical flow estimation using a spatial pyramid
  network.}},
\newblock \bibinfo{journal}{IEEE Computer Society Conference on Computer Vision
  and Pattern Recognition (CVPR)}  (\bibinfo{year}{2017}).
\bibitem[{Hui et~al.(2018)Hui, Tang, and Loy}]{Hui2018}
\bibinfo{author}{T.-W. Hui}, \bibinfo{author}{X.~Tang}, \bibinfo{author}{C.~C.
  Loy},
\newblock \bibinfo{title}{{LiteFlowNet: a lightweight convolutional neural
  network for optical flow estimation}},
\newblock \bibinfo{journal}{IEEE Computer Society Conference on Computer Vision
  and Pattern Recognition (CVPR)}  (\bibinfo{year}{2018}).
\bibitem[{Butler et~al.(2012)Butler, Wulff, Stanley, and Black}]{Butler2012}
\bibinfo{author}{D.~J. Butler}, \bibinfo{author}{J.~Wulff},
  \bibinfo{author}{G.~B. Stanley}, \bibinfo{author}{M.~J. Black},
\newblock \bibinfo{title}{A naturalistic open source movie for optical flow
  evaluation},
\newblock \bibinfo{journal}{European Conference on Computer Vision (ECCV)}
  (\bibinfo{year}{2012}).
\bibitem[{Marsza{\l}ek et~al.(2009)Marsza{\l}ek, Laptev, and
  Schmid}]{Hollywood2009}
\bibinfo{author}{M.~Marsza{\l}ek}, \bibinfo{author}{I.~Laptev},
  \bibinfo{author}{C.~Schmid},
\newblock \bibinfo{title}{{Actions in context}},
\newblock \bibinfo{journal}{IEEE Computer Society Conference on Computer Vision
  and Pattern Recognition (CVPR)}  (\bibinfo{year}{2009}).
\bibitem[{NVIDIA(2018{\natexlab{a}})}]{NVIDIA2018}
\bibinfo{author}{NVIDIA}, \bibinfo{title}{{NVIDIA open sources NVVL: library
  for machine learning training}}, \bibinfo{howpublished}{available at
  \url{https://hub.packtpub.com/nvidia-open-sources-nvvl-library-for-machine-learning-training/}},
  \bibinfo{year}{2018}{\natexlab{a}}.
\bibitem[{NVIDIA(2018{\natexlab{b}})}]{NVIDIAG2018}
\bibinfo{author}{NVIDIA}, \bibinfo{title}{{NVVL}},
  \bibinfo{howpublished}{available at \url{https://github.com/NVIDIA/nvvl}},
  \bibinfo{year}{2018}{\natexlab{b}}.
\bibitem[{Kingma and Ba(2014)}]{Kingma2014}
\bibinfo{author}{D.~P. Kingma}, \bibinfo{author}{J.~Ba},
\newblock \bibinfo{title}{{Adam: a method for stochastic optimization}},
\newblock \bibinfo{journal}{arXiv preprint arXiv:1412.6980}
  (\bibinfo{year}{2014}).
\bibitem[{Wang et~al.(2004)Wang, Bovik, {Rahim Sheikh}, and
  Simoncelli}]{Wang2004}
\bibinfo{author}{Z.~Wang}, \bibinfo{author}{A.~C. Bovik},
  \bibinfo{author}{H.~{Rahim Sheikh}}, \bibinfo{author}{E.~P. Simoncelli},
\newblock \bibinfo{title}{{Image quality assessment: from error visibility to
  structural similarity}},
\newblock \bibinfo{journal}{IEEE Transactions On Image Processing (TIP)}
  \bibinfo{volume}{13} (\bibinfo{year}{2004}).
\bibitem[{Netflix(2018{\natexlab{a}})}]{Netflix2018}
\bibinfo{author}{Netflix}, \bibinfo{title}{{VMAF: the journey continues}},
  \bibinfo{howpublished}{available at
  \url{https://medium.com/netflix-techblog/vmaf-the-journey-continues-44b51ee9ed12}},
  \bibinfo{year}{2018}{\natexlab{a}}.
\bibitem[{Netflix(2018{\natexlab{b}})}]{NetflixG2018}
\bibinfo{author}{Netflix}, \bibinfo{title}{{VMAF: video multi-method assessment
  fusion}}, \bibinfo{howpublished}{available at
  \url{https://github.com/Netflix/vmaf}}, \bibinfo{year}{2018}{\natexlab{b}}.
\bibitem[{University(2000)}]{VTL2000}
\bibinfo{author}{A.~S. University}, \bibinfo{title}{{Video trace library YUV
  video sequences}}, \bibinfo{howpublished}{available at
  \url{http://trace.kom.aau.dk/yuv/index.html}}, \bibinfo{year}{2000}.
\bibitem[{Kamble et~al.(2017)Kamble, Thakur, and Bajaj}]{Kamble2017}
\bibinfo{author}{S.~D. Kamble}, \bibinfo{author}{N.~V. Thakur},
  \bibinfo{author}{P.~R. Bajaj},
\newblock \bibinfo{title}{Modified three-step search block matching motion
  estimation and weighted finite automata based fractal video compression},
\newblock \bibinfo{journal}{International Journal of Interactive Multimedia and
  Artificial Intelligence (IJIMAI)}  (\bibinfo{year}{2017}).
\bibitem[{Yang et~al.(2010)Yang, Au, Pang, Dai, and Zou}]{Yang2010}
\bibinfo{author}{W.~Yang}, \bibinfo{author}{O.~C. Au},
  \bibinfo{author}{C.~Pang}, \bibinfo{author}{J.~Dai},
  \bibinfo{author}{F.~Zou},
\newblock \bibinfo{title}{{An efficient motion vector coding algorithm based on
  adaptive motion vector prediction}},
\newblock \bibinfo{journal}{IEEE International Symposium on Circuits and
  Systems (ISCAS)}  (\bibinfo{year}{2010}).

\end{thebibliography}
